\documentclass[a4paper,preprint,11pt,prd,nofootinbib,letterpaper,floatfix,superscriptaddress,preprintnumbers]{revtex4}
\pdfoutput=1

\usepackage[colorlinks=true
,urlcolor=blue
,anchorcolor=blue
,citecolor=blue
,filecolor=blue
,linkcolor=blue
,menucolor=blue
,pagecolor=blue
]{hyperref}
\usepackage[all]{hypcap}

\usepackage{color}

\usepackage{graphicx}

\usepackage{verbatim}
\usepackage{amsmath}
\usepackage{amssymb}
\usepackage[caption=false]{subfig}
\usepackage{url}

\usepackage{multirow}
\usepackage{threeparttable}
\usepackage{paralist}

\usepackage{color}
\definecolor{darkgreen}{rgb}{0,0.5,0}
\definecolor{darkred}{rgb}{0.7,0.0,0}
\definecolor{lightblue}{rgb}{0.0,0.5,1}

\newcommand{\SU}{\text{SU}}
\newcommand{\SO}{\text{SO}}

\newcommand{\EM}{\text{EM}}

\newcommand{\U}{\text{U}}

\renewcommand{\comment}[1]{}

\DeclareRobustCommand{\Sec}[1]{Sec.~\ref{#1}}
\DeclareRobustCommand{\Secs}[2]{Secs.~\ref{#1} and \ref{#2}}
\DeclareRobustCommand{\App}[1]{App.~\ref{#1}}
\DeclareRobustCommand{\Tab}[1]{Table~\ref{#1}}

\DeclareRobustCommand{\Fig}[1]{Fig.~\ref{#1}}

\DeclareRobustCommand{\Eq}[1]{Eq.~(\ref{#1})}
\DeclareRobustCommand{\Eqs}[2]{Eqs.~(\ref{#1}) and (\ref{#2})}
\DeclareRobustCommand{\Ref}[1]{Ref.~\cite{#1}}
\DeclareRobustCommand{\Refs}[1]{Refs.~\cite{#1}}

\newcommand{\be}{\begin{equation}}
\newcommand{\ee}{\end{equation}}

\newcommand{\beq}{\begin{equation}}
\newcommand{\eeq}{\end{equation}}

\newcommand{\mb}[1]{\boldsymbol{#1}}

\newcommand{\cL}{\mathcal{L}}

\newcommand{\cZ}{\mathcal{Z}}
\newcommand{\cO}{\mathcal{O}}

\begin{document}

\title{Colorful Twisted Top Partners and Partnerium at the LHC}

\author{Yevgeny Kats}
\email{yevgeny.kats@cern.ch}
\affiliation{Theoretical Physics Department, CERN, Geneva, Switzerland}
\affiliation{Department of Physics, Ben-Gurion University, Beer-Sheva 8410501, Israel}
\affiliation{Department of Particle Physics and Astrophysics, Weizmann Institute of Science, Rehovot 7610001, Israel}

\author{Matthew McCullough}
\email{matthew.mccullough@cern.ch}
\affiliation{Theoretical Physics Department, CERN, Geneva, Switzerland}

\author{Gilad~Perez}
\email{gilad.perez@weizmann.ac.il}
\affiliation{Department of Particle Physics and Astrophysics, Weizmann Institute of Science, Rehovot 7610001, Israel}

\author{Yotam Soreq}
\email{soreqy@mit.edu}
\affiliation{Center for Theoretical Physics, Massachusetts Institute of Technology,\\ Cambridge, MA 02139, USA}

\author{Jesse Thaler}
\email{jthaler@mit.edu}
\affiliation{Center for Theoretical Physics, Massachusetts Institute of Technology,\\ Cambridge, MA 02139, USA}

\date{\today}

\preprint{MIT-CTP {4897}}
\preprint{CERN-TH-2017-073}

\begin{abstract}
In scenarios that stabilize the electroweak scale, the top quark is typically accompanied by partner particles.
In this work, we demonstrate how extended stabilizing symmetries can yield scalar or fermionic top partners that transform as ordinary color triplets but carry exotic electric charges.
We refer to these scenarios as ``hypertwisted'' since they involve modifications to hypercharge in the top sector.
As proofs of principle, we construct two hypertwisted scenarios:  a supersymmetric construction with spin-0 top partners, and a composite Higgs construction with spin-1/2 top partners.
In both cases, the top partners are still phenomenologically compatible with the mass range motivated by weak-scale naturalness.
The phenomenology of hypertwisted scenarios is diverse, since the lifetimes and decay modes of the top partners are model dependent.
The novel coupling structure opens up search channels that do not typically arise in top-partner scenarios, such as pair production of top-plus-jet resonances.
Furthermore, hypertwisted top partners are typically sufficiently long lived to form ``top-partnerium'' bound states that decay predominantly via annihilation, motivating searches for rare narrow resonances with diboson decay modes.
\end{abstract}

\maketitle

\section{Introduction}
\label{sec:intro}

The discovery of the Higgs boson~\cite{Aad:2012tfa,Chatrchyan:2012xdj} not only cemented the structure of the standard model~(SM), but it also reemphasized the importance of symmetries (and symmetry breaking) for fundamental physics.
As the Large Hadron Collider~(LHC) continues to search for new phenomena, symmetries remain a useful guide for predicting possible extensions to the SM.
Of particular interest are symmetries---either exact or approximate---that relate SM particles to possible new partner states, since those symmetries could help stabilize the Higgs mass against quantum corrections and thereby resolve the hierarchy problem.
Since the top quark is responsible for the greatest sensitivity of the Higgs mass to physics at the cutoff through its Yukawa coupling, top partners are a ubiquitous prediction of beyond-the-SM scenarios and a key target for LHC searches.

In typical frameworks that address the hierarchy problem, including supersymmetry~(SUSY)~\cite{Dimopoulos:1981zb} and Higgs compositeness~\cite{Kaplan:1983fs,Kaplan:1983sm,ArkaniHamed:2002qx,ArkaniHamed:2002qy,Agashe:2004rs,Contino:2006qr}, the top partners often have the same color and electric charge as the top quark.
This occurs because the symmetry that stabilizes the Higgs potential commutes with the $\SU(3)_C \times \U(1)_{\EM}$ subgroup of the SM.
There are more exotic scenarios, however, where the charges of the top quark and top partner can differ, leading to unique LHC signatures.
For example, the top partners can be neutral under $\SU(3)_C$, yet still inherit the top quark's coupling to the Higgs boson due to a discrete or continuous symmetry.
These colorless top partners appear in models like twin Higgs~\cite{Chacko:2005pe,Chacko:2005vw,Chacko:2005un,Falkowski:2006qq,Chang:2006ra,Batra:2008jy,Craig:2013fga,Craig:2014aea,Craig:2014roa,Geller:2014kta,Barbieri:2015lqa,Low:2015nqa,Craig:2015pha,Craig:2016kue,Barbieri:2016zxn,Katz:2016wtw,Contino:2017moj,Badziak:2017syq}, quirky little Higgs~\cite{Cai:2008au}, and folded SUSY~\cite{Burdman:2006tz,Cohen:2015gaa}, and they could even play the role of dark matter~\cite{Poland:2008ev} or right-handed neutrinos~\cite{Batell:2015aha}.

In this paper, we explore the possibility of colorful twisted top partners, where the new states are still $\SU(3)_C$ triplets but carry exotic electric charges.
Such scenarios arise when the symmetry that stabilizes the electroweak scale is extended to include an exact or approximate $\mathcal{Z}_2$ symmetry that does not commute with $\U(1)_{\EM}$.
We refer to these scenarios as ``hypertwisted'' since the underlying mechanism involves modifying hypercharges in the top sector.
We provide example hypertwisted constructions both for spin-$0$ top partners arising from SUSY and for spin-$1/2$ top partners arising from Higgs compositeness.

By itself, the presence of top-like states with exotic electric charges is not so surprising, since the top partner could simply be part of a larger top multiplet with extended electroweak quantum numbers.
For example, composite Higgs scenarios often feature a color-triplet fermion with charge $5/3$ (see e.g.~\cite{DeSimone:2012fs}).
The key difference here is that the exotic state is a true top partner, in the sense that its radiative contribution to the Higgs potential cancels against the top quark, at least at one loop.

Exotic electric charges can lead to an accidental approximate $\mathcal{Z}_2$ symmetry, since charge conservation may prohibit any renormalizable couplings to the SM.
Colorful twisted top partners can therefore lead to rich phenomenology, since their decays to SM particles via higher-dimension operators will be model dependent.
They can be long-lived if they are the lightest new state carrying the accidental $\mathcal{Z}_2$, or they can be elusive due to decays to hadronic and/or multibody final states.
In addition, a potentially crucial signal for hypertwisted scenarios is ``partnerium'' production.
Since a pair of top partners carries no charge under the possible $\mathcal{Z}_2$ symmetry, top-partnerium bound states typically annihilate to pairs of gauge or Higgs bosons.
LHC diboson resonance searches therefore provide an important probe of such scenarios, particularly if the electric charge of the top partner is large, making the diphoton branching fraction sizable.
These bound states are the analogs of stoponium from SUSY, whose LHC signals (which appear much less generically than in the models we explore here) have been studied in~\Refs{Drees:1993yr,Drees:1993uw,Martin:2008sv,Martin:2009dj,Younkin:2009zn,Kumar:2014bca,Kats:2009bv,Kahawala:2011pc,Barger:2011jt,Kim:2014yaa,Batell:2015zla}.

As a historical note, the development of the SM already highlights a case where a partner particle required by naturalness was first discovered via partnerium.
In 1970, Glashow, Iliopoulos, and Maiani proposed that the up quark should have a generation-like partner---the charm quark---which was required to control the rate of strangeness-violating processes at the quantum level~\cite{Glashow:1970gm}.
The charm mass was predicted to be below 5\,GeV~\cite{Gaillard:1974hs}, but it was hard to observe in open channels due to a complicated set of off-shell-mediated final states (see e.g.~\cite{Appelquist:1978aq}).
Instead, the charmonium $J/\psi$ state was observed in 1974  at BNL~\cite{Aubert:1974js} and SLAC~\cite{Augustin:1974xw}, which fit well with the perturbative QCD postdiction~\cite{Appelquist:1974zd}.
One can envision a similar development for hypertwisted top partners, where top-partnerium could be discovered prior to open top-partner production at the LHC.

For the case of spin-$0$ top partners, our construction is related to SUSY in slow motion~\cite{Chacko:2008cb}, where the top partner and top quark share the same gauge quantum numbers, but are not directly part of the same ${\cal N}=1$ multiplet due to folded SUSY~\cite{Burdman:2006tz}.
Here, we both fold and hypertwist SUSY (to be distinguished from the twist in \Ref{Cohen:2015gaa}) to give the top partner an arbitrary electric charge.
For the case of spin-$1/2$ top partners, we explore hypertwisted composite Higgs models.
Analogous to the dark top scenario \cite{Poland:2008ev}, we introduce an enlarged global symmetry for the top multiplet and then use symmetry-violating mass terms to project out ordinary top partner states and retain only twisted states at low energies.

The remainder of this paper is organized as follows.  In \Sec{sec:ColorTwist}, we highlight the key ingredients for colorful twisted top partners and sketch the main phenomenological implications.  We then present two example constructions:  a SUSY scenario in \Sec{sec:scalar} and a composite Higgs scenario in \Sec{sec:fermionic}.  We explore several possibilities for top-partner decays and the accompanying top partnerium signals in \Sec{sec:pheno}, and conclude in \Sec{sec:summary}.  Additional details are provided in the appendices.

\section{Colorful twisted naturalness}
\label{sec:ColorTwist}

\begin{figure}
\centering
\subfloat[]{
\includegraphics{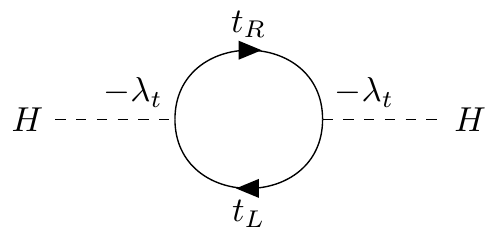} 
\label{fig:cancelling_top}
}
$\qquad$
\subfloat[]{
\includegraphics{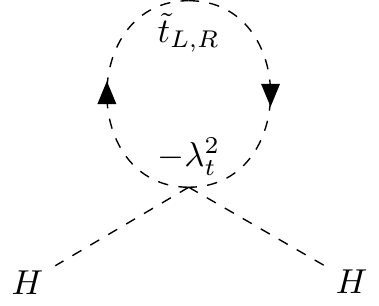} 
\label{fig:cancelling_spinzero}
}
$\qquad$
\subfloat[]{
\includegraphics{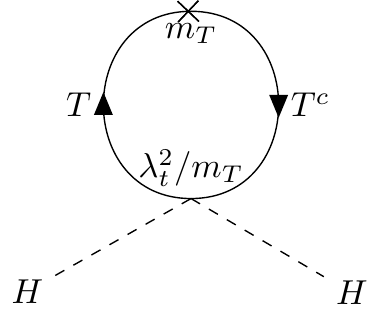} 
\label{fig:cancelling_spinhalf}
}
\caption{Minimal diagrams for the cancellation of Higgs quadratic divergences from the top Yukawa coupling.  (a)  The divergent SM top loop.  (b)  Cancellation through spin-0 top partners.  (c)  Cancellation through spin-1/2 top partners. }
\label{fig:cancelling}
\end{figure}

Let us briefly recap some key features of models that stabilise the electroweak scale using global symmetries.  At the level of one-loop Feynman diagrams, it is straightforward to determine the minimal structure needed to control radiative corrections to the Higgs potential from the large top Yukawa coupling.  The SM top Yukawa coupling, taken to be real for simplicity, is
\be
\label{eq:generic_SM}
\mathcal{L}_{\rm SM} \supset -\lambda_t q H t^c,
\ee
where $q$\,($t^c$) is the top electroweak doublet\,(singlet) and $H$ is the Higgs. \Eq{eq:generic_SM} leads to the famous quadratically divergent top-loop diagram in \Fig{fig:cancelling_top}.  For the case of spin-0 top partners, 
one has complex scalars, $\tilde{Q}_3$ and $\tilde{U}^c_3$, that get a contribution to their mass from electroweak symmetry breaking. For example, the interactions 
\be
\label{eq:generic_spinzero}
\mathcal{L}_\text{spin-0} \supset -m_{\tilde{Q}_3}^2 |\tilde{Q}_3|^2 - m_{\tilde{U}^c_3}^2 |\tilde{U}^c_3|^2 - \lambda_t^2 |H \cdot \tilde{Q}_3|^2 - \lambda_t^2 |H|^2 |\tilde{U}^c_3|^2
\ee
generate canceling diagrams shown in \Fig{fig:cancelling_spinzero}.  For the case of spin-1/2 top partners, $T$ and $T^c$, one has vector-like fermions whose mass, $m_T$, and Higgs coupling are correlated through a new scale $f = m_T/\lambda_t$. For example, in the limit $f \gg v$, the terms
\be
\label{eq:generic_spinhalf}
\mathcal{L}_\text{spin-1/2} \supset - \left(m_T - \frac{\lambda_t^2}{2 m_T}|H|^2 \right) T T^c
\ee
are sufficient to achieve the canceling diagram in \Fig{fig:cancelling_spinhalf}.  In both cases, each top partner state has to be a triplet, either of $\SU(3)_C$ or of a new global or gauged $\SU(3)$, in order to match the multiplicity of top states in the SM loop.  More general cancellation structures have been recently explored in \Ref{Craig:2014roa}.

At the two-loop level, diagrams with internal gauge bosons appear, so unless the top partners have the right gauge quantum numbers, there will be two-loop quadratic divergences.
From the perspective of a low-energy effective theory with cutoff $\Lambda$, though, these two-loop effects are subdominant and could be addressed in the corresponding ultraviolet~(UV) completion.  Therefore, the top partners need not carry color, as explored in the folded-SUSY/twin-Higgs literature~\cite{Chacko:2005pe,Chacko:2005vw,Chacko:2005un,Falkowski:2006qq,Chang:2006ra,Batra:2008jy,Craig:2013fga,Craig:2014aea,Craig:2014roa,Geller:2014kta,Barbieri:2015lqa,Low:2015nqa,Craig:2015pha,Craig:2016kue,Barbieri:2016zxn,Katz:2016wtw,Contino:2017moj,Badziak:2017syq,Burdman:2006tz,Cohen:2015gaa,Poland:2008ev,Batell:2015aha}.  Here, we focus on colorful top partners but exploit the freedom to hypertwist their electric charges away from $+2/3$.

An immediate consequence of these colorful twisted top partners is the presence of $\mathcal{Z}_2$ symmetries, which could be exact or approximate.  One $\mathcal{Z}_2$ symmetry, which we denote as $\mathcal{Z}^\lambda_2$, is needed to ensure that the exact same $\lambda_t$ coupling in \Eq{eq:generic_SM} appears also in \Eq{eq:generic_spinzero} or~\eqref{eq:generic_spinhalf}, otherwise the divergent pieces of the diagrams in \Fig{fig:cancelling} would not cancel.  In general, this $\mathcal{Z}^\lambda_2$ could be a subgroup of a larger symmetry.  We show proofs-of-concept that such $\mathcal{Z}^\lambda_2$ symmetries are possible in the hypertwisted constructions in \Secs{sec:scalar}{sec:fermionic}, where the field content of the SM is effectively doubled and then folded to project out unwanted states.

Another $\mathcal{Z}_2$ symmetry, which we denote as $\mathcal{Z}^T_2$, is more model dependent.  The interactions required for naturalness, in \Eqs{eq:generic_spinzero}{eq:generic_spinhalf}, respect a symmetry under which the top partners are odd. (The terms in these equations are actually invariant under a full $\U(1)$-partner symmetry, with the SM fields being neutral.)  This $\mathcal{Z}^T_2$ symmetry becomes an approximate symmetry of the whole theory (including the SM) if the charges of the partners are exotic enough to forbid low-dimension operators that would violate it. Consequently, exotic top partner charges often lead to partner longevity or even stability.
Note that for ordinary untwisted spin-1/2 top partners, this $\mathcal{Z}^T_2$ symmetry is not necessarily present, and in some cases the $T^c$ state would mix with the SM $t^c$; if one wants to suppress this mixing, an additional symmetry like $T$-parity~\cite{Cheng:2003ju,Cheng:2004yc,Low:2004xc} is required.  For hypertwisted spin-1/2 top partners with modified electric charges, though, $T^c$/$t^c$ mixing is forbidden, leading to the approximate $\mathcal{Z}^T_2$-symmetric structure.  An interesting exception is when the hypertwisted top partner has the same quantum numbers as the bottom quark, in which case there is no $\mathcal{Z}^T_2$ symmetry since $T^c$/$b^c$ mixing is allowed. To simplify the discussion, we will not pursue that possibility in the present work, though we note that the resulting phenomenology is expected to be similar to the  ``Beautiful Mirrors" model~\cite{Choudhury:2001hs}.

Since stable colored particles are excluded up to high masses (e.g.\ 1.2~TeV for a color-triplet scalar with charge $2/3$~\cite{CMS-PAS-EXO-16-036,Aaboud:2016uth}), a light top partner must be able to decay.  
However, if there is an exact $\mathcal{Z}^T_2$ symmetry, then the lightest $\mathcal{Z}^T_2$-odd particle could be color-neutral (e.g.\ a hypertwisted lepton, gauge boson, or Higgs boson partner).  Let us refer to this as the LZP.
Some high-scale interaction at the cutoff $\Lambda$ could mediate the decay of the top partner to the LZP, for example through an off-shell massive gauge boson (as in the case of charm decay).
If the LZP is electrically neutral, it could be a dark matter candidate, and the top partner electric charge is then fixed by the specific decay mode of the top partner to the LZP.
In this dark matter case, top partner decays would face bounds from standard SUSY searches.

Alternatively, this $\mathcal{Z}^T_2$ might only be approximate, in which case the top partner could be the LZP and decay to SM particles.  The electric charge of the top partner is then constrained by the availability of decay modes, which in turn restricts the electric charge of the top partner to be an integer difference from $2/3$.
In the case of a scalar top partner, the decay can be to two quarks and/or leptons, similar to $R$-parity-violating~(RPV) stop decays in SUSY~\cite{Barbier:2004ez}.
For fermionic top partners, two-body dipole transitions or three-body decays are both possible.  For decays that involve final-state leptons or neutrinos, there are rather stringent bounds from the LHC; light top partners are only possible assuming mostly hadronic decays.  If the $\mathcal{Z}^T_2$ is only broken by $\Lambda$-suppressed interactions, then twisted top partner LZPs are expected to be considerably longer-lived than ordinary top partners.  In this way, top partners could exhibit displaced decays, a feature also present in SUSY-in-slow-motion scenarios~\cite{Chacko:2008cb}.

Regardless of whether the $\mathcal{Z}^T_2$ symmetry is exact or approximate, a potential important prediction of hypertwisted top partners is the presence of near-threshold QCD bound states of top partner pairs.
Unlike in ordinary untwisted cases, where the constituent decays typically dominate over bound-state annihilation (similar to the SM toponium) or even prevent bound-state formation altogether (if the decay rate is larger than the binding energy), the suppressed decay rate of the twisted top partners (due to the approximate $\mathcal{Z}^T_2$) preserves the bound-state annihilation signals. In cases where the top partners have elusive decays, partnerium annihilation could be the dominant signal of colorful twisted naturalness, as discussed further in \Sec{sec:pheno}.

\section{Spin-0 example:  hyperfolded SUSY}
\label{sec:scalar}

In this section, we present an explicit model using the techniques of folded SUSY~\cite{Burdman:2006tz,Cohen:2015gaa} to establish a theory of scalar colored top partners which carry an arbitrary electric charge.  
This setup uses an exact exchange symmetry to robustly enforce the $\mathcal{Z}^\lambda_2$ required for the top partner to regulate the one-loop Higgs potential.
We call this ``hyperfolded SUSY'', since hypercharge, rather than color, participates in the folded SUSY construction.\footnote{An equally apt name would be  ``hypertwisted SUSY'', though twisting has another meaning in the SUSY context~\cite{Cohen:2015gaa}.}
We refer to states with ordinary quantum numbers as SM states, 
while those with exotic quantum numbers as hyperfolded states.

We first consider the structure in the UV: a SUSY theory in 5D.
The gauge and matter multiplets live in the bulk with $\mathcal{N}=1$ SUSY in 5D.
From the 4D perspective, the matter fields live in $\mathcal{N}=2$ hypermultiplets with vector-like field content, which can be written as pairs of $\mathcal{N}=1$ chiral multiplets.
At the compactification scale, SUSY is broken via the Scherk-Schwarz mechanism~\cite{Scherk:1979zr,Scherk:1978ta,Cremmer:1979uq,Fayet:1985ua,Fayet:1985kt,Rohm:1983aq,Kounnas:1988ye,Ferrara:1988jx,Kounnas:1989dk,Kiritsis:1996xd}.
The extra dimension $y$ is compactified to $S^1/\mathcal{Z}_2$, with fixed points at $y=0$ and $y = \pi R$, and SUSY breaking arises due to boundary conditions at these fixed points.
While the structure of the theory can be understood in terms of $R$ and orbifold symmetries, we find it more pragmatic to simply discuss the boundary conditions on individual component fields.\footnote{If required, these boundary conditions can be derived through representations under discrete subgroups of the 5D symmetries.}

Let us consider the quark superfields first.  The discussion follows~\Ref{Chacko:2008cb}.
The $\mathcal{N}=2$ quark hypermultiplet contains two Weyl fermions $\psi_Q,\psi_Q^c$ and two complex scalars $\widetilde{Q},\widetilde{Q}^c$.  
There are two different ways to organize these fields into 4D $\mathcal{N}=1$ chiral multiplets, either by pairing $\mb{Q} = (\psi_Q,\widetilde{Q})$ (and similarly for the conjugate fields), or by pairing $\mb{Q}' = (\psi_Q,\widetilde{Q}^{c\ast})$.
On the orbifold boundaries we can choose to constrain the fields with boundary conditions.  As in folded SUSY, we conserve the first kind of $\mathcal{N}=1$ SUSY at $y=0$ and the second kind of $\mathcal{N}=1$ SUSY at $y=\pi R$, via the component-field boundary conditions
\be
\psi_Q (+,+),\quad \psi_Q^c (-,-),\quad \widetilde{Q}(+,-),\quad \widetilde{Q}^c (-,+),
\label{eq:bcmat}
\ee
where we have selected to have propagating~($+$) or constrained~($-$) component fields at the boundaries~($0,\pi R$).
This leaves only the fermion $\psi_Q$ zero mode as a propagating field at low energies.  In this way, the boundary conditions have removed the zero modes of three out of four fields contained within the $\mathcal{N}=2$ quark hypermultiplet.

Analogous to folded SUSY, our construction contains a hyperfolded copy of the quark superfields with modified hypercharge.  We indicate the hyperfolded sector fields with an $F$ subscript.  We impose the boundary conditions
\be
\psi_{Q_F} (+,-),\quad \psi_{Q_F}^c (-,+) ,\quad \widetilde{Q}_F(+,+),\quad \widetilde{Q}_F^c (-,-),
\label{eq:bcfol}
\ee
which leaves only the scalar $\widetilde{Q}_F$ zero mode at low energies.  For the gauge hypermultiplets, which contain the adjoint vector $A^a_{\mu}$, two Weyl fermions $\lambda^a, \lambda^{a^c}$, and one complex scalar $\widetilde{\sigma}^a$, we impose the boundary conditions
\be
A^a_{\mu} (+,+),\quad \lambda^a (-,+),\quad \lambda^{a^c} (+,-),\quad \widetilde{\sigma}^a (-,-),
\ee
to leave only the gauge fields at low energies.  To summarize, for each boundary in isolation there is a full $\mathcal{N}=1$ SUSY, but this SUSY is not the same at each boundary, having been flipped amongst the hypermultiplet members.   This leaves only one field out of each hypermultiplet at low energies.

\begin{figure}
\centering
\includegraphics[width=0.6\textwidth]{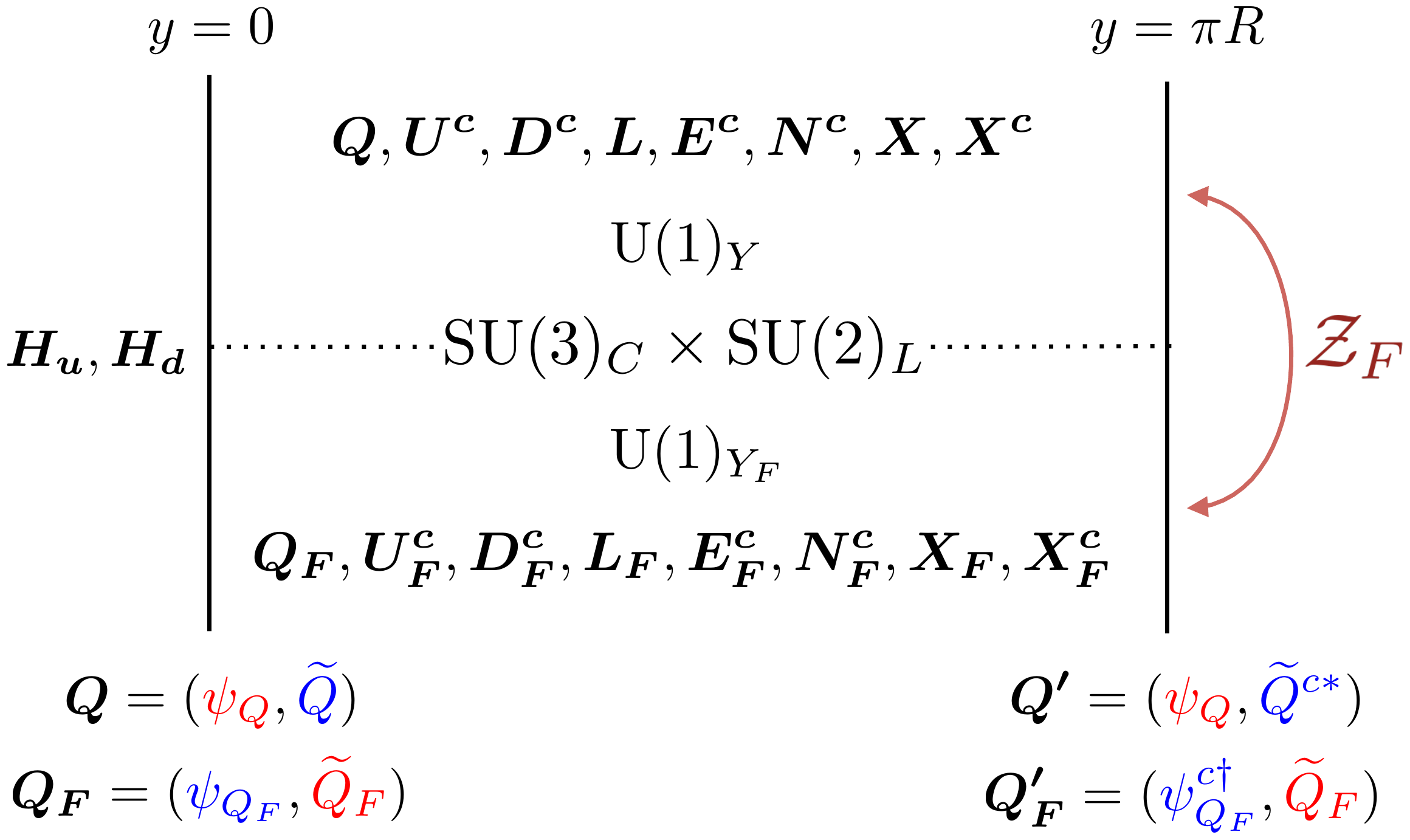} 
\caption{Illustration of the hyperfolded SUSY model. The fields in red have a zero mode while the fields in blue do not.  Note that the conjugate fields do not have propagating zero modes. 
}
\label{fig:folded}
\end{figure}

We must also decide where to put the Higgs multiplets.  As we will see, the equality of the Yukawa couplings will be enforced by an exchange symmetry $\mathcal{Z}_F$ between the SM and hyperfolded sectors, $\mb{Q} \leftrightarrow \mb{Q_F}$\,.  Thus, the Higgs multiplets must live at an orbifold point in which the boundary conditions respect $\mathcal{Z}_F$.  By comparing \Eq{eq:bcmat} and \Eq{eq:bcfol} we see that the only such point is $y=0$, thus we place the full Higgs chiral multiplets and Yukawa couplings at $y=0$.   A schematic illustration of this construction is given in \Fig{fig:folded}.\footnote{Note that the SUSY-breaking one-loop contributions to the Higgs mass parameter from the top and folded top superfields follow the usual Scherk-Schwarz pattern, described in detail in \Ref{ArkaniHamed:2001mi}.  Importantly, as matter and folded matter have opposite twist parameters, these one-loop contributions, which can be thought of as containing the usual top/stop contributions, cancel.  This cancellation does not persist at two loops.}

The complete matter content and gauge representations of the model are given in \Tab{tab:matter} using $\mathcal{N}=1$ language.
The key new ingredient is a new gauge group $\U(1)_{Y_F}$ which participates in the $\mb{Q} \leftrightarrow \mb{Q_F}$ exchange symmetry and allows us to achieve the hyperfolded charge assigments.
The $\U(1)_{Y_F}$ gauge charges are proportional to a linear combination of hypercharge and $\U(1)_{B-L}$, and right-handed neutrinos $\mb{N^c}$ have been added, such that the low energy field content is anomaly free.  More specifically, we have set
\beq
Y_F = Y + (3q-2)(B-L) \,,
\eeq
where the coefficient of the first term has to be $1$ for the Yukawa interactions to preserve both $\U(1)_Y$ and $\U(1)_{Y_F}$, while the coefficient of the second term is a free parameter, which we have written in terms of the resulting electric charge $q$ of the hyperfolded stops.
To avoid exactly stable top partners (or, more generally, stable charged particles if the top partner is not the lightest new state), $q - 2/3$ must be taken to be an integer, which is an additional model-building assumption.

The relevant Yukawa terms in the superpotential at $y = 0$ are given by
\begin{eqnarray}
\mb{W}_{\rm Yuk} & = &  \lambda_u \mb{H_u} \left(\mb{Q} \mb{U^c}+\mb{Q_F} \mb{U_F^c}  \right) -\lambda_d \mb{H_d} \left(\mb{Q} \mb{D^c}+\mb{Q_F} \mb{D_F^c}  \right) \nonumber\\
& & -\lambda_e \mb{H_d} \left(\mb{L} \mb{E^c}+\mb{L_F} \mb{E_F^c}  \right) +\lambda_\nu \mb{H_u} \left(\mb{L} \mb{N^c}+\mb{L_F} \mb{N_F^c}  \right)\,.
\label{eq:sup}
\end{eqnarray}
As in \Ref{Chacko:2008cb}, one may additionally have the usual $\mu$ term in the Higgs sector and also add NMSSM-like Higgs singlet couplings to raise the Higgs mass and generate the appropriate $B_\mu$ terms.

\newcommand{\doublecharge}[2]{#1 \leftrightarrow #2}

\begin{table}[t]
\centering
\begin{tabular}{r @{ $\leftrightarrow$ } l @{$\quad$} c @{$\quad$}c @{$\quad$} r @{ $\leftrightarrow$ } l @{$\qquad$} r @{ $\leftrightarrow$ } l}
\hline\hline
\multicolumn{2}{c@{$\quad$}}{} & $\SU(3)_{C}$ & $\SU(2)_L$ & \multicolumn{2}{c@{$\qquad$}}{$\U(1)_Y$} & \multicolumn{2}{c}{$\U(1)_{Y_F}$} \\
\hline \hline
\multicolumn{2}{c@{$\quad$}}{$\mb{H_u}$} & $\mb{1}$ & $\mb{2}$ & \multicolumn{2}{c@{$\qquad$}}{$1/2$} & \multicolumn{2}{c}{$1/2$} \\
\multicolumn{2}{c@{$\quad$}}{$\mb{H_d}$} & $\mb{1}$ & $\mb{2}$ & \multicolumn{2}{c@{$\qquad$}}{$-1/2$} & \multicolumn{2}{c}{$-1/2$} \\
\hline
$\mb{Q}$&$\mb{Q_F}$ & $\mb{3}$ & $\mb{2}$ & $\frac{1}{6}$&$q-\frac{1}{2}$ & $q-\frac{1}{2}$&$\frac{1}{6}$ \\
$\mb{U^c}$&$\mb{U^c_F}$ & $\mb{\overline{3}}$ & $\mb{1}$ & $-\frac{2}{3}$&$-q$ & $-q$&$-\frac{2}{3}$ \\
$\mb{D^c}$&$\mb{D^c_F}$ & $\mb{\overline{3}}$ & $\mb{1}$ & $\frac{1}{3}$&$1-q$ & $1-q$&$\frac{1}{3}$ \\
$\mb{L}$&$\mb{L_F}$ & $\mb{1}$ & $\mb{2}$ & $-\frac{1}{2}$&$\frac{3}{2}-3q$ & $\frac{3}{2}-3q$&$-\frac{1}{2}$ \\
$\mb{E^c}$&$\mb{E^c_F}$ & $\mb{1}$ & $\mb{1}$ & $1$&$3q-1$ & $3q-1$&$1$ \\
$\mb{N^c}$&$\mb{N^c_F}$ & $\mb{1}$ & $\mb{1}$ & $0$&$3q-2$ & $3q-2$&$0$ \\ 
$\mb{X}$&$\mb{X_F}$ & $\mb{1}$ & $\mb{1}$  & $q_X$&$0$ & $0$&$q_X$ \\
$\mb{X^c}$&$\mb{X^c_F}$ & $\mb{1}$ & $\mb{1}$  & $-q_X$&$0$ & $0$&$-q_X$\\
\hline\hline
\end{tabular}
\caption{The chiral matter content and gauge representations of the hyperfolded SUSY model, where the $F$ subscript indicates fields in the hyperfolded sector.  The exact exchange symmetry $\mathcal{Z}_F$ swaps the SM matter superfields for the hyperfolded matter superfields (i.e.\ $\mb{Q} \leftrightarrow \mb{Q_F}$) and the $\U(1)_Y$ and $\U(1)_{Y_F}$ gauge bosons.  The $\SU(3)_{C}$ and $\SU(2)_L$ gauge fields are unchanged under the exchange symmetry.  The $\mb{X}$ fields are introduced as a proxy for $\U(1)_{Y_F}$ breaking. We do not show the additional chiral multiplets which, along with the fields shown, complete the $\mathcal{N}=2$ hypermultiplets at the compactification scale.  }
\label{tab:matter}
\end{table}

The equality of the original and hyperfolded superfield couplings to the Higgs boson is enforced by the $\mathcal{Z}_F$ exchange symmetry described in \Tab{tab:matter} and illustrated in \Fig{fig:folded}.  The only states remaining below the compactification scale are the known SM fermions (with the neutrinos being Dirac), the gauge fields, the hyperfolded scalars, and the Higgs bosons and higgsinos.  Most importantly for naturalness, the largest couplings between the Higgs and matter fields are given by 
\begin{eqnarray}
	\label{eq:LSUSY}
	\mathcal{L} 
	& \supset &
	\lambda_t H_u \psi_{Q_3} \psi_{U^c_3} + \lambda_b H_d \psi_{Q_3} \psi_{D^c_3} 
	- \lambda_t^2 \left( |H_u \cdot \widetilde{Q}_{3F}|^2 + |H_u|^2 |\widetilde{U}^c_{3F}|^2 \right) \nonumber\\
	& & - \lambda_b^2 \left( |H_d \cdot \widetilde{Q}_{3F}|^2 + |H_d|^2 |\widetilde{D}^c_{3F}|^2 \right)+ \ldots,
\end{eqnarray}
which is precisely of the form in \Eq{eq:generic_spinzero}, demonstrating that the third-generation hyperfolded stop and sbottom squarks, which may be light, play the role of the top and bottom partners.  The ellipsis denotes additional terms less relevant for naturalness.

We must also consider the hyperfolded $\U(1)_{Y_F}$ gauge symmetry which was introduced to complete the $\mb{Q} \leftrightarrow \mb{Q_F}$ exchange symmetry.  Clearly the associated gauge boson, $B_F$, would have been observed if it were light, thus we must somehow remove it from the spectrum.  If one simply removed this gauge symmetry by hand, the exchange symmetry would be broken and the equality of couplings in \Eq{eq:sup}, at least at the compactification scale, would become questionable.  This is not an insignificant point, because hypercharge contributions to supersymmetric wavefunction renormalization would in general lead to different values of couplings in \Eq{eq:sup}.\footnote{Similarly, we could remove the $B_F$ boson with a boundary condition at $y=\pi R$.   One should be careful, however, with other brane localized terms that might spoil the $\mathcal{Z}_F$ symmetry.}  To justify the equality of the couplings, we instead break the $\U(1)_{Y_F}$ gauge symmetry via the Higgs mechanism, introducing new superfields $\mb{X_F}$, $\mb{X_F^c}$ (and untwisted partners $\mb{X}$, $\mb{X^c}$).  We assume the SUSY-breaking soft terms for the scalar components of $\mb{X_F}$ and $\mb{X_F^c}$ are such that $\U(1)_{Y_F}$ is spontaneously broken and $B_F$ is sufficiently heavy to avoid limits on $Z'$ resonances from the LHC.  As the analogous $\mb{X}$ and $\mb{X^c}$ fields do not have a tachyonic soft mass (otherwise they would break hypercharge), this setup breaks the exchange symmetry, but only softly, thus it does not damage the radiative stability of the theory.  The hypercharged fermions in $\mb{X},\mb{X^c}$ can have vector-like masses from a $\mu$ term, thus they may be at or well above the weak scale.  
{Note that $B_F$ can be given a few-TeV mass without affecting the naturalness of the model, since the Higgs mass sensitivity to $m_{B_F}$ scales as $\delta m^2_H \sim g^2_Y m_{B_F}^2 /16\pi^2$. This contribution is comparable in size to the standard Bino contribution in the MSSM, which results in a mild contribution to the Higgs mass tuning, see e.g.~\cite{Papucci:2011wy}. Moreover, this effect is subdominant to the fact that, as in~\Refs{Burdman:2006tz,Chacko:2008cb}, the gauge boson loop contributions in our model do not get canceled until the scale $1/R$, which can be $\sim 10$~TeV, while $B_F$ can be lighter without contradicting current LHC bounds.}

The charge $q_X$ was intentionally left as a free parameter, to allow a variety of decay scenarios for the hyperfolded top partners.  After $\U(1)_{Y_F}$ breaking, the only remaining gauge symmetries are the SM gauge symmetries.  This means that if we wish for a hyperfolded scalar to decay via a particular operator $\mb{\mathcal{O}_F}$ that respects the SM gauge symmetries but carries non-zero $\U(1)_{Y_F}$ charge, we may use the operator $\mb{X_F} \mb{\mathcal{O}_F}$ with appropriate choice of $q_X$, which will typically be non-renormalizable.  This addition is not central to the model, but is rather a module by which we can study the general phenomenology of hyperfolded scenarios more fully.

Before considering specific phenomenological features, it is worthwhile to consider the broad features of this class of models.  Let us begin with the hyperfolded squark sector, in particular, its flavor structure. 
The simplest structure is obtained when the first two generations also live in the bulk and have the same boundary conditions as the third generation. In this flavor-blind case, the only source of explicit flavor breaking is coming from the Yukawa couplings, hence this setup belongs to the minimal-flavor-violating class of models~\cite{DAmbrosio:2002vsn,Kagan:2009bn}.   
As for the spectrum, the hyperfolded squarks, being part of incomplete chiral multiplets, receive finite contributions to their masses~\cite{Burdman:2006tz}: universal contributions from the gauge interactions, and non-universal ones from the Yukawa interactions. 
Consequently, the hyperfolded stop masses are of order of  $0.1/R$~\cite{Burdman:2006tz}, and about 20\% heavier than 
the first two generation squarks~\cite{Chacko:2008cb,Cohen:2015gaa}. 

Another notable relevant feature of hyperfolded SUSY is the absence of gauginos in the low-energy spectrum.  While naturalness arguments would require usual Majorana gluinos to show up at a few TeV, such a requirement does not arise in our setup.  As the full theory becomes $\mathcal{N}=2$ SUSY at the compactification scale, many of the desirable features of Dirac gauginos arise (see e.g.~\cite{Fox:2002bu,Kribs:2012gx}). In particular, naturalness only requires the first gluino Kaluza-Klein (KK) mode to be below roughly $5$~TeV, corresponding to inverse compactification scale, $1/R \sim 10$~TeV~\cite{Chacko:2008cb}.
The Dirac nature of the gluinos also implies that the squark production processes do not benefit from the valence-quark enhancement of the cross section, which leads to a weaker bound on their masses~\cite{Heikinheimo:2011fk,Kribs:2012gx,Mahbubani:2012qq}.
For example, six quark flavors decaying to dijet pairs could be as light as $\sim 800$~GeV~\cite{ATLAS-CONF-2016-084}, while more complicated mostly-hadronic decays would likely be undetected even for lower masses. 

An alternative flavor structure is to choose boundary conditions for the first two generations such that only the visible sector quarks, and none of the first two generation fields in the hyperfolded sector, remain below the compactification scale.  In this way, the only light colored scalars would be the hyperfolded stops and sbottoms, while all other colored scalars live at $m\sim 1/R$.  Due to the small Yukawas this would not impact the naturalness of the setup, yet it would remove the additional colored states beyond collider bounds.
This setup, however, now consists of two a priori unrelated sources of flavor breaking, the boundary conditions and the Yukawa interactions. Thus, in order not to generate overly large contributions to flavor-changing processes, a microscopic alignment mechanism is implicitly assumed to be active in such a case.  This is not a severe problem, though, since it could be achieved in a UV theory that possesses some form of flavor symmetry.
Another potential worry for our setup is that higher-dimensional operators, which generically mix different flavors and are suppressed only by the cutoff of the 5D theory, may lead to too large contributions to Kaon mixing and CP violation.
This is a standard issue for effective 5D theories with low cutoffs and, also in this case, flavor symmetries can lead to sufficient suppression and compatibility with constraints (see for instance~\cite{Gedalia:2010rj,Isidori:2010kg,KerenZur:2012fr} and references therein).

Because the Higgs multiplets live at $y= 0$, the higgsinos remain in the low energy spectrum, as also expected from naturalness.  This is because, like in the MSSM, their mass is given by the $\mu$ term which enters the Higgs potential.  That said, a light higgsino will not generically be involved in the hyperfolded squark decays.  In ordinary folded SUSY, the higgsino does not couple the folded squarks to SM quarks, but rather to the folded quarks which do not have zero modes.  Similarly, in hyperfolded SUSY, the hyperfolded squarks can decay to the higgsino only through more complicated processes whose rate can easily be suppressed relative to direct decays to SM particles induced by higher-dimension operators, even when the couplings responsible for the latter are relatively small.  For this reason, we neglect higgsinos in our later discussion of hyperfolded phenomenology in \Sec{sec:pheno}.

\section{Spin-1/2 example:  hypertwisted composite Higgs}
\label{sec:fermionic}

In this section, we sketch an example of a hypertwisted composite Higgs model, which demonstrates the possibility of having spin-1/2 top partners with arbitrary electric charges.
This toy model is based on standard composite Higgs ideas but with an enlarged global symmetry group, leading to the general features described in \Sec{sec:ColorTwist}.
It also shares some features with Little Higgs models (see~e.g.~\cite{Schmaltz:2005ky,Perelstein:2005ka} for a review), but without collective symmetry breaking for the Higgs quartic coupling. 
To ensure the cancellation of the top loop by the partner loop, the global symmetry group contains a $\cZ^\lambda_2$ symmetry that relates the top Yukawa in \Eq{eq:generic_SM} to the di-Higgs coupling of the partner in \Eq{eq:generic_spinhalf}.
Moreover, the model has a $\cZ^T_2$ symmetry acting on the partner which, in combination with an accidental symmetry due to the partner's exotic charge, generically suppresses partner decays.

To simplify the presentation, the toy model below is based on the coset space $\SU(3)/\SU(2)$.  This coset space does not exhibit custodial protection, so it is likely in conflict with electroweak precision tests.
In \App{app:SO5SO4}, we present a hypertwisted version of the minimal custodial-protected composite model based on the coset space $\SO(5)/ \SO(4)$~\cite{Agashe:2004rs}.
One could also consider constructions based on the twin Higgs mechanism~\cite{Chacko:2005pe}, where instead of an enlarged global symmetry (e.g.~$\SU(2)_F$ in the construction below), the $\cZ^\lambda_2$ symmetry is implemented directly on the top partners.
Compared to \Sec{sec:scalar}, we present fewer details on the possible UV completion and do not discuss the flavor structure at all, though we note that many of the challenges of constructing realistic UV embedding are shared with the composite Higgs literature (see e.g.~\cite{Panico:2015jxa} for a recent review). 
Moreover, we do not attempt to construct a realistic Higgs potential. 

We begin with a global symmetry
\be
\label{eq:SU3SU2symmter}
\SU(3)_G \times \SU(2)_F \times {\U(1)_Z} \,,
\ee
where the $F$ subscript is a reminder that the matrix
\be
\label{eq:foldingopertation}
\exp \left[ i \pi T^2_F \right]  = \left(\begin{array}{cc}0 & 1 \\ -1 & 0\end{array}\right)_F
\ee
performs an analogous folding operation to the $\mb{Q} \leftrightarrow \mb{Q_F}$ exchange symmetry from \Sec{sec:scalar}.\footnote{To make the analogy more precise, one can lift $\SU(2)_F$ to $\U(2)_F$ and use the matrix $\left(\begin{array}{cc}0 & 1 \\ 1 & 0\end{array}\right)$.}
We then introduce a (linear) sigma field $\Phi$ that transforms under $(\SU(3)_G, \SU(2)_F)_{\U(1)_Z}$ as:
\be
(\boldsymbol{\bar{3}},\boldsymbol{1})_{\frac{1}{3}}: \quad \Phi \,.
\ee
When $\Phi$ obtains a vacuum expectation value (vev), the symmetry breaking pattern is
\begin{align}
	\SU(3)_G \times \U(1)_Z 
	\ \to \ 
	\SU(2) \times \U(1) \,,
\end{align}
with $\SU(2)_F$ unaffected.

Expanding around the vev, the $\Phi$ field takes the form
\begin{align}
	\label{eq:Phi}
	\Phi = \exp\left[ - i \frac{\pi^a T^a_G}{f} \right]
	\begin{pmatrix}
		0 \\ 0 \\ f
	\end{pmatrix}
	\supset
	\begin{pmatrix}
	H \\
	f - \frac{H^\dagger H}{2f}
	\end{pmatrix} \, ,
\end{align}
where $\pi^a$ are the Goldstone modes, $T^a_G$ with $a=4,\ldots,8$ are the broken $\SU(3)_G$ generators, and $f$ is the symmetry breaking scale.  In the last step of \Eq{eq:Phi}, we have identified the SM Higgs as
\begin{align}
	H 
= 	-\frac{1}{2}\begin{pmatrix}
	\pi^5 + i \, \pi^4 \\
	\pi^7 + i \, \pi^6
	\end{pmatrix} \, \Rightarrow \begin{pmatrix} 0 \\ v/\sqrt2 \end{pmatrix} ,
\end{align}
where $v \approx 246$~GeV, and we expand in $H/f$ to second order.  The interactions below will respect the $T^8_G$ generator, such that $\pi^8$ is an exact Goldstone mode that only has derivative couplings.  Because $\pi^8$ can be decoupled from the spectrum either by introducing a soft mass or by gauging the $T^8_G$ generator, we do not consider $\pi^8$ in our analysis below for simplicity. 

The SM electroweak gauge group, preserved by the vev of $\Phi$, is identified with the following generators which are weakly gauged:
\begin{align}
	T_L^{1,2,3} = T_G^{1,2,3}\,, \qquad\qquad
	Y = Z - \frac{T_G^8}{\sqrt3} + \left(\frac23 - y_T\right)T^3_F \,,
\end{align}
where $y_T$ is a free parameter that will become the hypercharge (and electric charge) of the top partner of interest.  Similar to the hyperfolded SUSY case, one has to assume that $y_T - 2/3$ is an integer to avoid stable top partners (or other stable charged states).  Note that the hypercharge generator $Y$ does not commute with the $T_F^2$ generator in \Eq{eq:foldingopertation}.  As in \Sec{sec:scalar}, we must rely on the structure of the UV completion to ensure that the hypercharge contribution to wavefunction renormalization does not spoil the coupling structure in \Eqs{eq:generic_SM}{eq:generic_spinhalf}.  This occurs, for example, in holographic composite Higgs completions, where the $\SU(2)_F$ corresponds to a bulk gauge symmetry broken to hypercharge via a brane-localized Higgs mechanism \cite{Contino:2003ve,Thaler:2005en}.

Focusing on the top sector, the relevant matter content is
\begin{align}
	\label{eq:QQccomp}
	(\boldsymbol{3},\boldsymbol{2})_{\frac{y_T}{2}}: \quad Q = \begin{pmatrix}
	b & q'_d \\
	-t & -q'_u \\
	t' & T
	\end{pmatrix} \, , \qquad
	(\boldsymbol{1},\boldsymbol{\bar 2})_{-\frac{y_T}{2}-\frac{1}{3}}: \quad Q^c = \begin{pmatrix}
	t^c & -T^c
	\end{pmatrix} \,,
\end{align}
where the third generation SM doublet is $q \equiv (t,b)$ and there is an extra electroweak doublet $q' \equiv (q'_u,q'_d)$.  The top partner of interest is associated with $T$ and $T^c$, while $t'$ and $q'$ can be decoupled from the low-energy spectrum, as discussed below.
The SM charges of the various fields in this model are summarized in \Tab{tab:CompositeSM}.

\begin{table}[t]
\begin{center}
\begin{tabular}{c @{$\quad$} c @{$\quad$} c @{$\quad$} c}
\hline\hline
  & $\SU(3)_C$ & $\SU(2)_L$ & $\U(1)_Y$ \\
  \hline \hline
 $H$   & $\boldsymbol{1}$         & $\boldsymbol{2}$ & $1/2$ \\\hline
 $q$   & $\boldsymbol{3}$         & $\boldsymbol{2}$ & $1/6$  \\
 $t^c$ & $\boldsymbol{\bar{3}}$   & $\boldsymbol{1}$ & $-2/3$ \\\hline
 $T$   & $\boldsymbol{3}$         & $\boldsymbol{1}$ & $y_T$  \\
 $T^c$ & $\boldsymbol{\bar{3}}$   & $\boldsymbol{1}$ & $-y_T$ \\\hline
 $q'$  & $\boldsymbol{3}$         & $\boldsymbol{2}$ & $y_T - 1/2$ \\
 $q^{\prime c}$ & $\boldsymbol{\bar{3}}$ & $\boldsymbol{2}$ & $-(y_T - 1/2)$ \\
 $t'$  & $\boldsymbol{3}$         & $\boldsymbol{1}$ & $2/3$ \\
 $t^{\prime c}$ & $\boldsymbol{\bar{3}}$ & $\boldsymbol{1}$ & $-2/3$ \\
 \hline\hline
\end{tabular}
\caption{The SM quantum numbers of the different fields of the hypertwisted composite Higgs model.  The hypertwisted top partner is $T$/$T^c$, whereas the primed fields, which are needed to form complete $\SU(3)_G$ multiplets, can be pushed to the cutoff.}
\label{tab:CompositeSM}
\end{center}
\end{table}%

The Yukawa interaction, which contains the SM top Yukawa, is
\begin{align}
	\label{eq:LY}
	\cL_{Y}
=	\lambda_t Q  \Phi Q^c + \mbox{h.c.}\,,
\end{align}
where in the limit $f \gg v$, $\lambda_t$ is the SM top Yukawa coupling.  To achieve a hypertwisted low-energy spectrum, we want to decouple the states denoted with primes, $t'$ and $q'$. One possibility is through soft breaking terms of the global symmetry, giving the primed states vectorlike masses with new fields $t^{\prime c}$ and $q^{\prime c}$,
\begin{align}
	\label{eq:Lsoft}
	\cL_{\rm soft}
=	-M_{t'} t' t^{\prime c} - M_{q'} q' q^{\prime c} + \mbox{h.c.} \, .
\end{align}
This occurs, for example, in extra-dimensional setups with an $\SU(3)_G$ gauge symmetry in the bulk, where the zero modes of the unwanted fields are projected out by boundary conditions, as discussed further below.  
The interaction term of \Eq{eq:LY} leads to the following couplings between the Higgs boson and the SM and partner fermions:
\begin{align}
	\label{eq:LYexpand}
	\cL_{Y} \supset
	-\lambda_t q H t^c - \lambda_t \left(f - \frac{H^\dagger H}{2f}\right) T T^c 
	+\lambda_t q' H T^c + \lambda_t \left(f - \frac{H^\dagger H}{2f}\right) t' t^c 
	 + \cO(1/f^2) \, ,
\end{align}
which matches to \Eqs{eq:generic_SM}{eq:generic_spinhalf}. 

Next, we discuss the masses of the various fermions and show that the Higgs potential is free of quadratic divergences from fermion loops. Combining \Eqs{eq:LY}{eq:Lsoft}, one can write the fermion mass terms as
\begin{align}
	\cL_{\rm mass}
=	-\begin{pmatrix} t & t' \end{pmatrix} M_{2/3}  \begin{pmatrix} t^c \\ t^{\prime c} \end{pmatrix}
	-\begin{pmatrix} T & q'_u \end{pmatrix} M_{y_T} \begin{pmatrix} T^c \\ q^{\prime c}_u \end{pmatrix} + \mbox{h.c.}\,,
\end{align}
where $M_{2/3}\;(M_{y_T})$ is the mass matrix of the $Q_{\rm EM}=2/3\;(y_T)$ fermions, given by
\begin{align}
	M_{2/3} 
=	\begin{pmatrix}
		\lambda_t f s_\epsilon & 0 \\
		-\lambda_t f c_\epsilon & M_{t'}
	\end{pmatrix} , \qquad\quad
	M_{y_T} 
=	\begin{pmatrix}
		\lambda_t f c_\epsilon & 0 \\
		-\lambda_t f s_\epsilon & M_{q'}
	\end{pmatrix} \, ,
	\label{eq:Mfermion}
\end{align}
where
\be
s_\epsilon \equiv \sin\epsilon, \qquad c_\epsilon \equiv \cos\epsilon, \qquad \epsilon \equiv \frac{v}{\sqrt{2}f}.
\ee 
The Coleman-Weinberg potential for the Higgs~\cite{Coleman:1973jx} can then be computed as
\begin{align}
	V(\epsilon) 
=	-\frac{1}{32\pi^2}{\rm tr}\left[ M^\dagger M \Lambda^2 \right] 
	+\frac{1}{32\pi^2}{\rm tr}\left[ \left( M^\dagger M \right)^2 \log \left( \frac{\Lambda^2}{M^\dagger M} \right) \right] \, ,
\end{align}
where $M\equiv M(\epsilon)$ is the combination of the mass matrices from Eq.~\eqref{eq:Mfermion} and $\Lambda$ is the UV cutoff scale.
As long as the trace of the fermion mass-squared matrix is independent of $\epsilon$, the Higgs mass is free of quadratic divergences, at least at one loop.
We find that (disregarding the color multiplicity)
\begin{align}
	\label{eq:TrM23MYT}
	\mbox{tr}[M_{2/3}^\dagger M_{2/3}] = \lambda_t^2 f^2 + M_{t'}^2 \,, \quad\quad 
	\mbox{tr}[M_{y_T}^\dagger M_{y_T}] = \lambda_t^2 f^2 + M_{q'}^2 \, .
\end{align}
In general, this spectrum results in logarithmic divergences in the Higgs potential, which are proportional to $\mbox{tr}[ (M^\dagger M)^2]$.  Interestingly, the limit $M_{t'} = M_{q'}$ has an enhanced pseudo-$\SU(3)_G$ symmetry such that the one-loop Coleman-Weinberg potential is zero from the top sector.  Of course, this pseudo-$\SU(3)_G$ symmetry does not persist at the two-loop level since $t'$ and $q'$ have different electric charges.  

In the above setup, the top partners arose from the complete multiplets presented in \Eq{eq:QQccomp}, and the primed fermions, $q'$ and $t'$, were made heavy due to soft-breaking mass terms which paired them with vector-like partners.
This type of soft breaking is well-motivated in the context of extra-dimensional scenarios, where the fermions live in a 5D bulk (potentially with warped geometry), see for example~\cite{ArkaniHamed:1999dc,Gherghetta:2000qt,Huber:2000ie,Cai:2008au}.
If we assume $\SU(3)_G$-preserving Higgs and Yukawa interactions localized on one brane, then we must choose appropriate boundary conditions for the fermions on the other brane.
In 4D language, a bulk fermion contains both left- and right-handed components, thus it is vector-like.
With Dirichlet boundary conditions on the other brane, a specific chirality can be projected out of the theory, only to appear as a heavy state of mass $m\sim1/R$.
Similarly, with Neumann boundary conditions on the other brane, a chiral zero mode will persist.
To achieve the mass terms in \Eq{eq:Lsoft}, we can therefore choose Dirichlet boundary conditions for both chiralities of the 5D fermions $q'_d, q'_u, t'$, such that none of these modes survive as zero modes in the theory.
To achieve the 4D chiral zero modes, we can choose Neumann boundary conditions for one chirality of $b,t,T$, and Dirichlet for the other chirality, furnishing the desired top sector and top partners to participate in \Eq{eq:QQccomp}.
As long as the fermions have Neumann boundary conditions on the $\SU(3)_G$-preserving brane, violations of $\SU(3)_G$ will be suppressed by the inter-brane separation~\cite{Contino:2003ve,Thaler:2005en}.

Note that in these extra-dimensional constructions, there are KK top partners with the standard hypercharge assignment.  While these KK modes have the expected couplings to regulate the top contribution to the Higgs potential, it is misleading to think of them as true top partners.  Instead, their radiative corrections largely balance against those from the KK hypertwisted top partners, such that quadratic divergences in the Coleman-Weinberg potential cancel KK level by KK level.  The primary role of these top quark KK modes is to regulate the residual logarithmic divergences away from the $M_{t'} = M_{q'}$ limit; their mass can therefore be significantly higher than the electroweak scale.  That said, the KK scale cannot be much higher than $1/R \sim$ 5--10 TeV, since this sets the mass of the KK gauge bosons, which do behave like partner particles to regulate the $W$/$Z$ boson contributions to the Higgs potential.

As discussed in \Sec{sec:ColorTwist}, one can identify an approximate $\cZ^T_2$ symmetry, under which all the particles with electric charge of $y_T$ are odd and the ones with charge of 2/3 and all the SM particles are even.\footnote{In the case where $y_T = -1/3$, this $\cZ^T_2$ symmetry may be spoiled by mixing between the top partner and down-type quarks.  The expected size of this mixing term depends on how exactly the down-type singlet quark is embedding into an $\SU(2)_F$ multiplet.}  To ensure that the lightest $\cZ^T_2$-odd particle is not stable, we assume the presence of additional higher-dimensional interactions that mediate top-partner decays, as we now discuss in \Sec{sec:pheno} below.

\section{LHC signatures}
\label{sec:pheno}

Having set the stage for colored top partner states with exotic electric charges, we now discuss their collider phenomenology.
We start our discussion by analyzing the resonant signals from annihilation of partneria, near-threshold QCD bound states of top partner pairs.
As mentioned already, these signals are generically present in the hypertwisted scenarios due to the (approximate) $\cZ^T_2$ symmetry. 
While the cross section for partnerium production is typically smaller than the cross section for continuum top partner pair production, the partnerium signatures are very clean (especially in the diphoton channel).
Moreover, the signals are independent of the top partner decays, as long as the decay is not too fast.

We then turn to top partner pair production signatures.
Because of the considerable freedom in the gauge quantum numbers of the top partners, as well as freedom in the masses and couplings of other particles that may be involved in top partner decays, there is an enormous range of phenomenological possibilities.
Indeed, even within a single framework, such as the MSSM, where the top partner properties are fixed, there are diverse possibilities for top partner decays.
For this reason, we do not attempt an exhaustive study of the different possibilities, but only present several model-dependent examples, focusing on cases which are different from standard scenarios and whose coverage by existing searches might be suboptimal.

\subsection{Top partnerium}

Possible LHC signals of QCD bound states of particles with exotic electric charges have been studied systematically in~\Ref{Kats:2012ym}, and more recently in~\Refs{Han:2016pab,Kats:2016kuz,Blum:2016szr}. Via gauge interactions alone, spin-0 $S$-wave bound states of such particles can be produced from gluon fusion and annihilate to $gg$, $\gamma\gamma$, $\gamma Z$, $ZZ$, and $W^+W^-$ (if the particles are charged under $\SU(2)_L$).

Importantly, the studies in~\Refs{Kats:2012ym,Han:2016pab,Kats:2016kuz,Blum:2016szr} assumed the particle couplings to the Higgs to be negligible relative to their couplings to gluons. For top partners, though, this assumption is not satisfied since $y_t \sim g_s$.  As discussed below, the couplings of top partners to the Higgs may lead to large annihilation rates to pairs of $W$/$Z$/Higgs bosons, and correspondingly reduced branching fractions to other (e.g.\ diphoton) final states.

In the hyperfolded SUSY model of \Sec{sec:scalar}, the partners are color-triplet scalars with an arbitrary electric charge. The partner of interest could be either the right-handed stop, $\widetilde{U}^c_{3F}$, or the upper or lower component of the left-handed doublet, $\widetilde{Q}_{3F}$. The partner's coupling to the Higgs, from the third term in \Eq{eq:LSUSY},\footnote{There is also a contribution from $D$ terms, which shifts the coupling as
$$
\lambda^2_t \to \lambda^2_t + \frac12\, g_Y^2 \cos 2\beta \times \left(Q + \frac23\,,\; \frac{T_3}{\sin^2\theta_W} - Q - \frac16\right)
$$
for the singlet and doublet, respectively, where $g_Y = 2m_Z\sin\theta_W/v$ and we approximated the $\U(1)_{Y_F}$ gauge coupling by $g_Y$. For $|Q| \leq 5/3$, the $D$ terms shift $\lambda_t^2$ by an amount between $-22\%$ (for a left-handed stop with $Q=-4/3$) and $+26\%$ (for a left-handed sbottom with $Q=5/3$), where to maximize the effect we have assumed $\tan\beta \gg 1$. For definiteness, this is also the limit we will assume in our plots.} can produce large partial annihilation widths to $WW$, $ZZ$, and $hh$. In the case of $\widetilde{U}^c_{3F}$, all the three modes will be important, while in the case of $\widetilde{Q}_{3F}$ the stop will have large rates to $ZZ$ and $hh$, and the sbottom to $WW$.  For some ranges of parameters, annihilation to $t\bar t$ is sizable as well. The expressions are similar to those obtained for the MSSM stoponium (see e.g.~the appendix of~\cite{Martin:2008sv}).\footnote{For the Higgs self coupling, which is present in one of the diagrams contributing to the $hh$ rate, we take the SM value, even though ${\cal O}(1)$ deviations are possible (see e.g.~\cite{Wu:2015nba}).} As analyzed in \App{impact-on-production}, the enhancement in binding due to Higgs exchange is negligible.  On the other hand, the reduction in the diphoton branching fraction from the $WW$, $ZZ$, $hh$, and $t\bar t$ decay channels has to be taken into account when interpreting limits.

\begin{figure}[t]
\begin{center}
\includegraphics[width=0.45\textwidth]{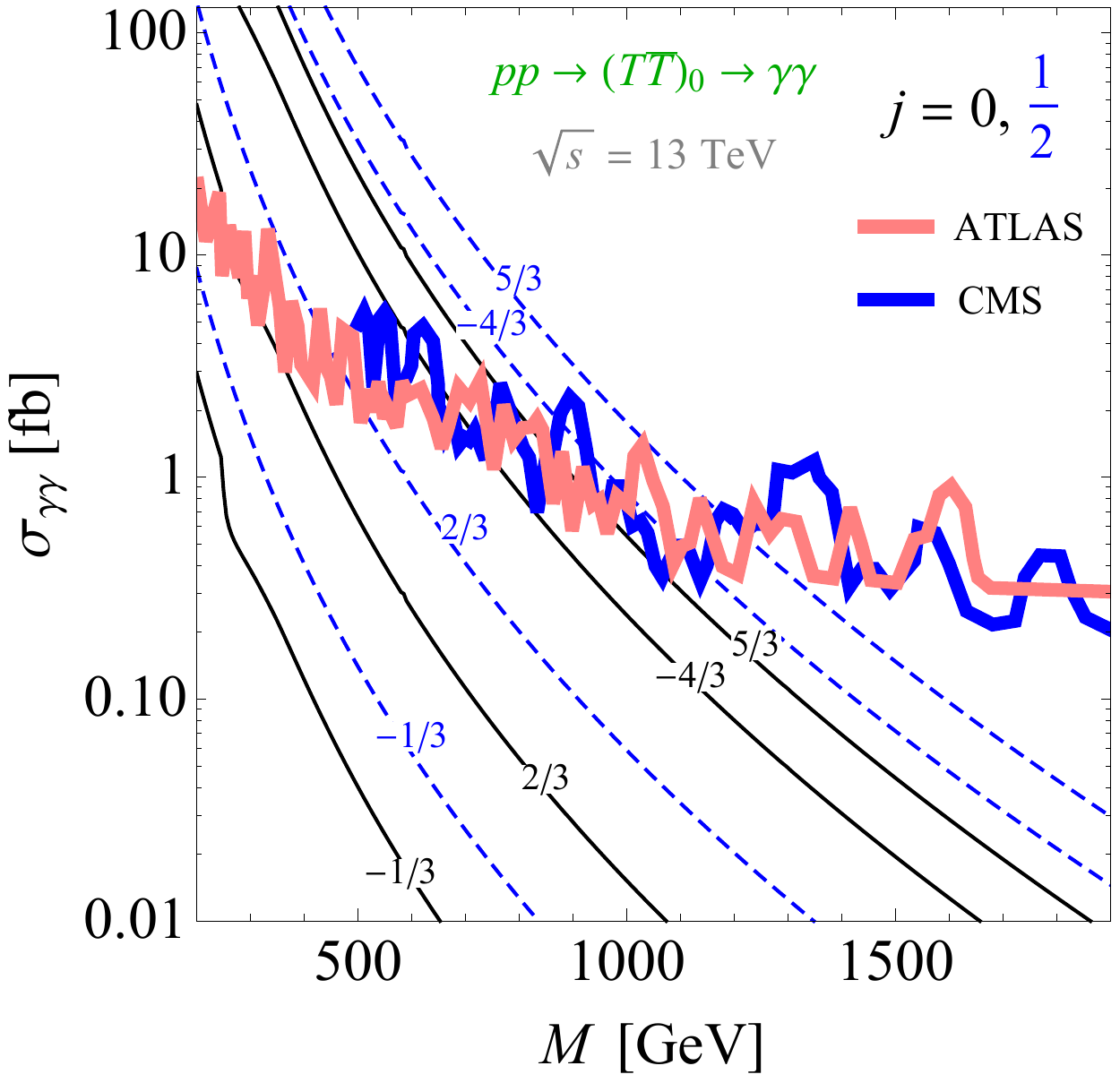}\qquad\quad
\includegraphics[width=0.45\textwidth]{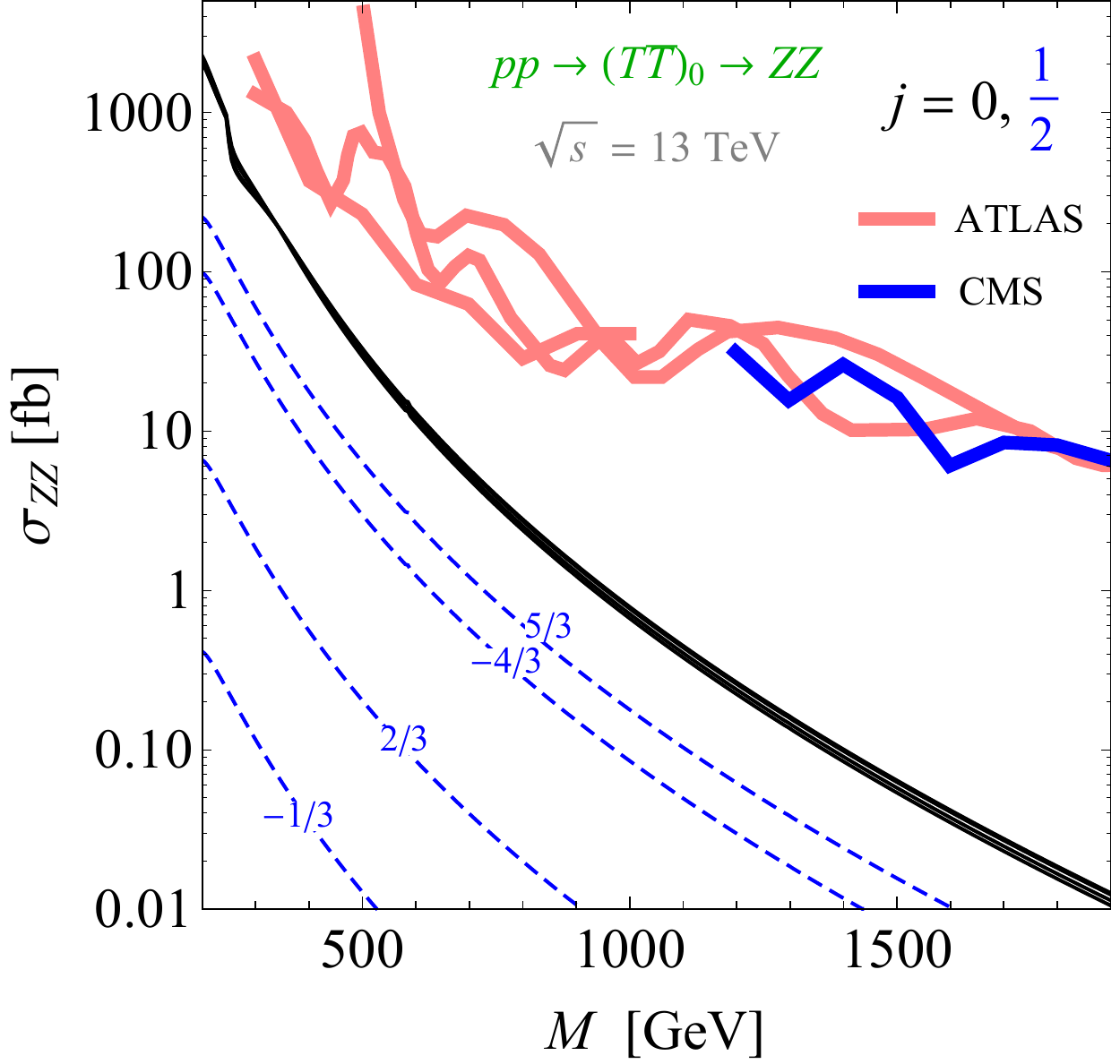}\\\vspace{5mm}
\includegraphics[width=0.45\textwidth]{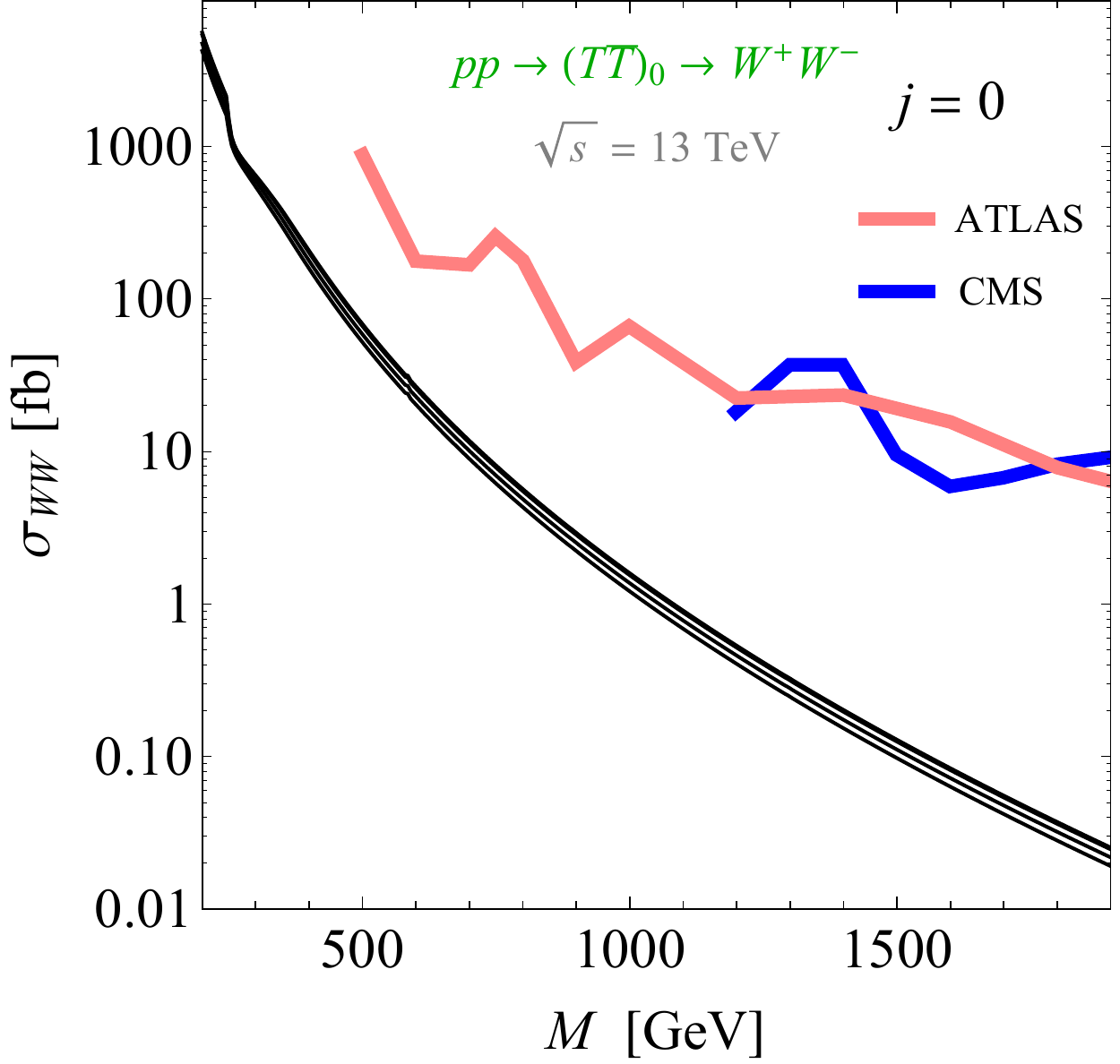}\qquad\quad
\includegraphics[width=0.45\textwidth]{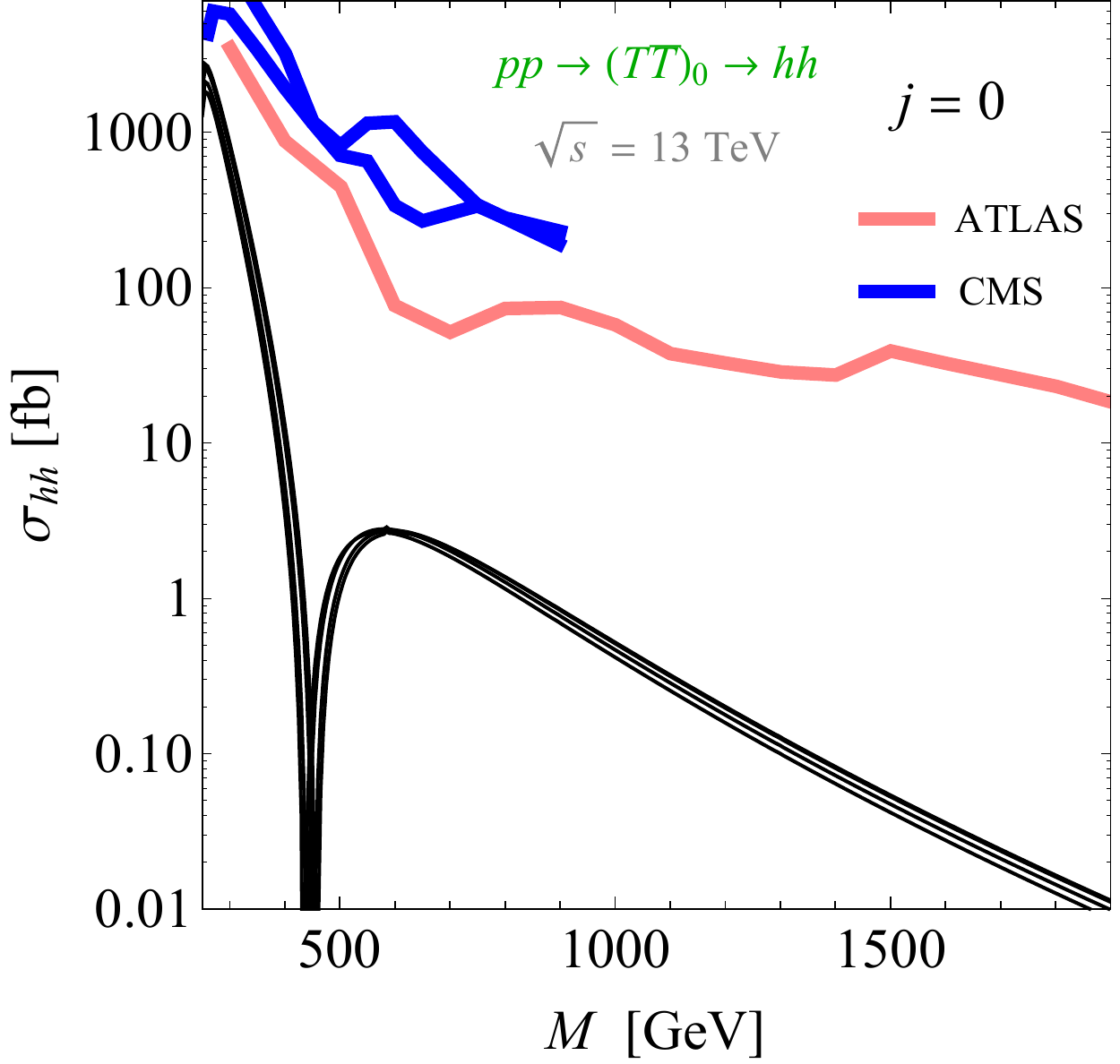}\vspace{-2mm}
\end{center}
\caption{Spin-0 partnerium signals at the 13 TeV LHC in the (top-left) $\gamma\gamma$, (top-right) $ZZ$, (bottom-left) $WW$, and (bottom-right) $hh$ channels.  
Shown are the cross sections for $\SU(2)_L$-singlet scalars (solid black) and fermions (dashed blue) for electric charge values indicated on each curve, as a function of the partnerium mass $M$.  In the $ZZ$, $WW$, and $hh$ channels, the curves for scalars are very close to each other because these rates are dominated by the Higgs coupling. There are no $WW$ or $hh$ modes for $\SU(2)_L$-singlet fermions.  The rates are subject to an overall QCD uncertainty of roughly a factor of $2$, as discussed in~\Ref{Kats:2016kuz}.  Also shown are the latest LHC limits on resonances decaying to $\gamma\gamma$ (ATLAS, 15~fb$^{-1}$~\cite{ATLAS-CONF-2016-059}; CMS, 13~fb$^{-1}$~\cite{CMS-PAS-EXO-16-027}), $ZZ$ (ATLAS, 13~fb$^{-1}$~\cite{ATLAS-CONF-2016-056,ATLAS-CONF-2016-082}; CMS, 36~fb$^{-1}$~\cite{CMS-PAS-B2G-17-001}), $WW$ (ATLAS, 13~fb$^{-1}$~\cite{ATLAS-CONF-2016-062}; CMS, 36~fb$^{-1}$~\cite{CMS-PAS-B2G-17-001}) and $hh$ (ATLAS, 13~fb$^{-1}$~\cite{ATLAS-CONF-2016-049}; CMS, 36~fb$^{-1}$~\cite{CMS-PAS-HIG-17-002,CMS-PAS-HIG-17-006}).  The $j$ in the legend refers to the spin of the top partner.}
\label{fig-limits}
\end{figure}

In \Fig{fig-limits}, we show the predicted signal cross section and current ATLAS~\cite{ATLAS-CONF-2016-059,ATLAS-CONF-2016-056,ATLAS-CONF-2016-082,ATLAS-CONF-2016-062,ATLAS-CONF-2016-049} and CMS~\cite{CMS-PAS-EXO-16-027,CMS-PAS-B2G-17-001,CMS-PAS-HIG-17-002,CMS-PAS-HIG-17-006} limits for the case of an $\SU(2)_L$ singlet with several choices for the electric charge.
We use the leading-order MSTW2008 parton distribution functions~\cite{Martin:2009iq} and, based on the results of~\Refs{Martin:2009dj,Younkin:2009zn}, apply an approximate $K$ factor of $1.4$ to the $gg$ production and annihilation rates.  The wavefunction at the origin is treated as in~\Refs{Kats:2012ym,Kats:2016kuz}, and contributes an overall uncertainty of roughly a factor of $2$ to the rates shown in \Fig{fig-limits}, as discussed in~\Ref{Kats:2016kuz}. Since for scalar constituents the decays to $WW$, $ZZ$, and $hh$ are dominated by the operator of Eq.~\eqref{eq:generic_spinzero}, their rates are almost independent of the electric charge chosen, hence the four curves corresponding to the different electric charges are practically on top of each other for these channels.
The dip in the $hh$ plot is due to a cancellation between the four contributing diagrams (contact interaction, $s$-channel higgs, and $t$- and $u$-channel stop).

In \Sec{sec:fermionic}, we presented a toy model in which a fermionic top partner can have an arbitrary electric charge. 
In this case, the Higgs coupling, from the second term in  \Eq{eq:LYexpand}, does not lead to any new or enhanced annihilation modes for the spin-0 $S$-wave bound state. Indeed, for fermionic constituents this bound state is a pseudoscalar, so cannot annihilate to $hh$ or pairs of longitudinal $W$ or $Z$ bosons. By explicit calculation, we find that the leading-order $Zh$ decay mode is vanishing as well.
As a result, different from the scalar case, the signals (also shown in \Fig{fig-limits}) are the same as without the Higgs coupling.\footnote{While precision electroweak constraints on the modified Higgs couplings typically require $v/f \lesssim 1/3$ (see e.g.\ \cite{DeSimone:2012fs,Grojean:2013qca}), and therefore top partner masses $m \sim \lambda_t f \gtrsim 750$~GeV, we include results for lower masses for completeness.}

For fermionic top partners, one should also consider the spin-1 $S$-wave bound state, which is absent in the scalar case. Despite the non-negligible dilepton branching fraction of this bound state (via an $s$-channel $Z/\gamma$), the signal is not necessarily easy to see because resonant QCD production of this state, from either the $gg$ or $q\bar q$ initial state, is impossible. Instead, as studied in~\Ref{Kats:2012ym}, there are contributions from resonant electroweak production from $q\bar q$, production from $gg$ in association with a $g$, $\gamma$, or $Z$, and deexcitation of $gg$-produced $P$-wave states. Production from $gg$ in association with the Higgs is forbidden by charge conjugation invariance (as known for SM quarkonia~\cite{Kniehl:2002wd,Kniehl:2004fa}).

\begin{figure}[t]
\begin{center}
\includegraphics[width=0.48\textwidth]{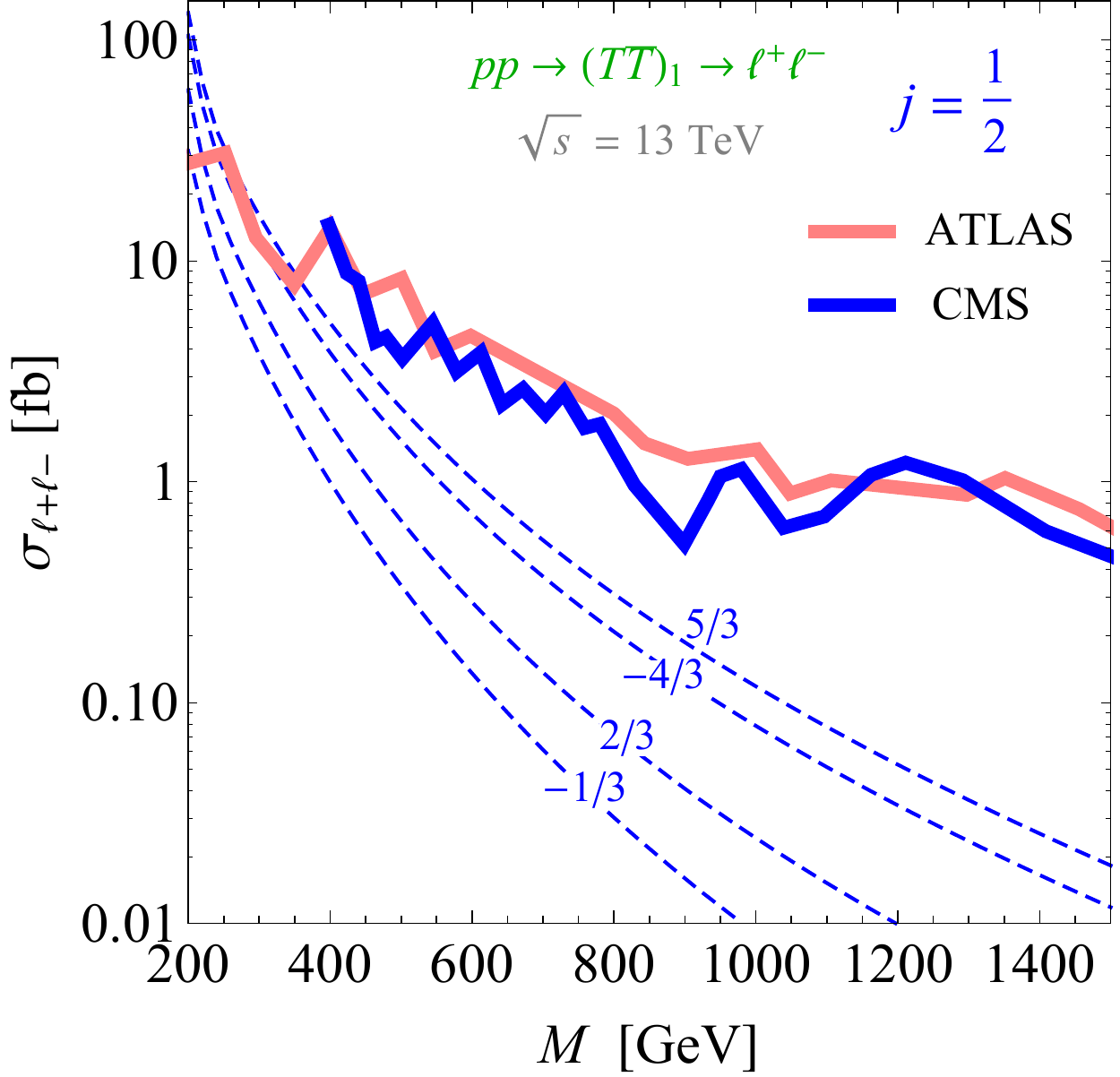}\quad
\includegraphics[width=0.48\textwidth]{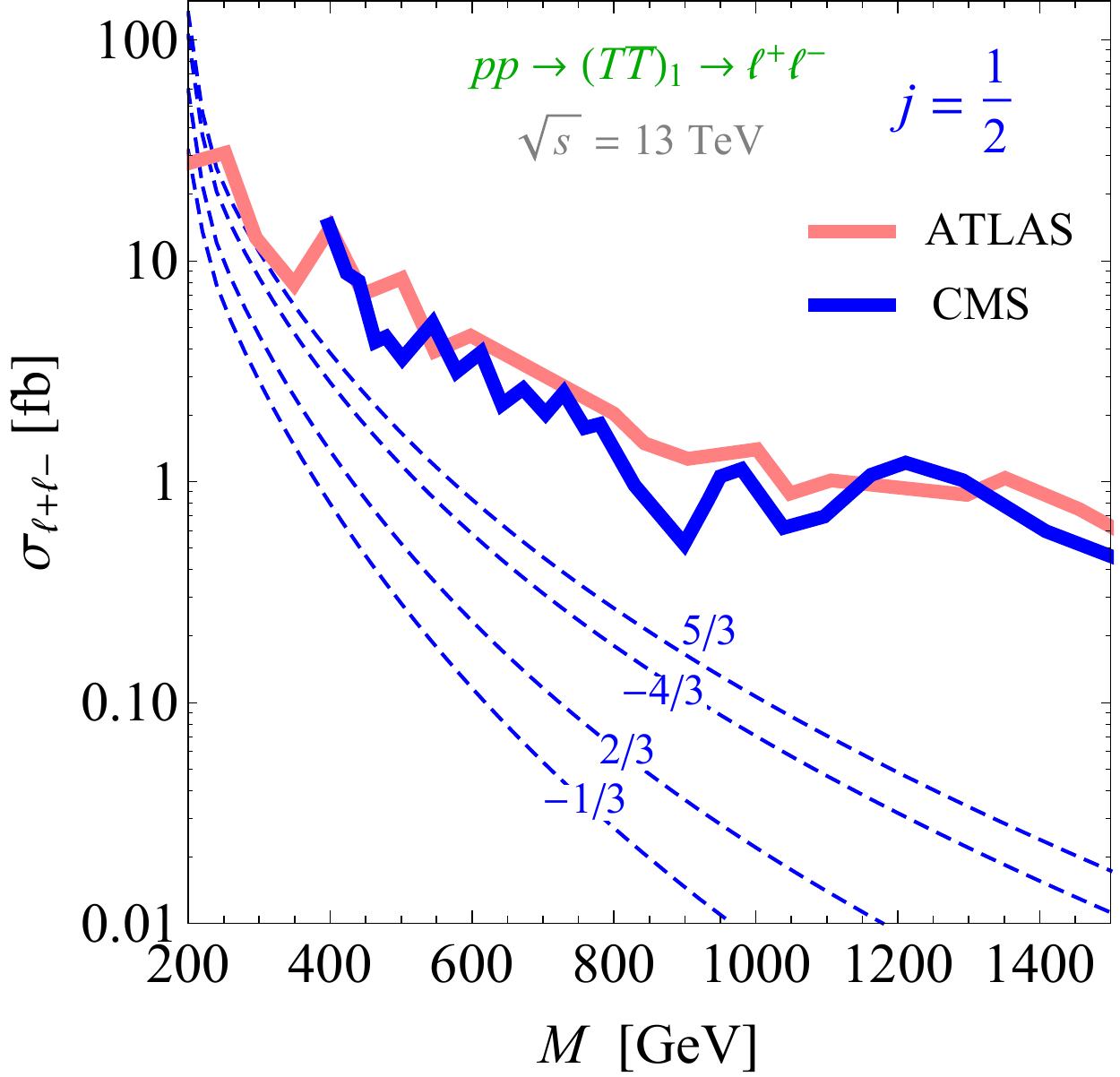}
\vspace{-2mm}
\end{center}
\caption{Spin-1 partnerium signals at the 13 TeV LHC in the $\ell^+\ell^-$ channel (for any single  flavor of leptons), including the branching ratio suppression due to the $Zh$ mode, for the Higgs coupling of \Eq{eq:generic_spinhalf} (left) or \Eq{eq:LYexpand54} (right).  The signal cross sections (dashed blue) are shown for electric charge values indicated on each curve, as a function of the partnerium mass $M$.  As in \Fig{fig-limits}, the rates are subject to an overall QCD uncertainty of roughly a factor of $2$.  Also shown are the latest LHC limits ($\simeq 13$~fb$^{-1}$) on $\ell^+\ell^-$ resonances from ATLAS~\cite{ATLAS-CONF-2016-045} and CMS~\cite{CMS-PAS-EXO-16-031}.  Note that these plots only hold for $j = 1/2$ top partners.}
\label{fig-limits-dilepton}
\end{figure}

The spin-1 $S$-wave bound state has a suppressed dilepton branching fraction in the presence of the Higgs coupling due to an enhanced annihilation rate to $Zh$.
When the partner mass $m \gg v$, the $Zh$ rate is independent of the Higgs coupling and given by
\beq
\label{eq:GammaZh}
\left. \Gamma_{(T\bar T)_1 \to Zh} \right|_{m \gg v} \simeq \frac{Q^2\alpha\tan^2\theta_W\,m_Z^2}{4v^2m^2}\,|\psi(\mathbf{0})|^2 \,.
\eeq
where $\psi(\mathbf{0})$ is the bound state wave function at the origin. When $m \sim {\cal O}(v)$, however, the rate is significantly enhanced and becomes comparable to the total annihilation rate to fermion pairs, thus leading to a non-negligible reduction of the dilepton signal relative to the case without the Higgs coupling. For example, for $m=300\,$GeV, $\Gamma_{(T\bar T)_1 \to Zh}$ is larger by a factor of 13\,(37) than \Eq{eq:GammaZh} for the model of \Sec{sec:fermionic}\,(\App{app:SO5SO4}), while for $m=1\,$TeV the enhancement is only a factor of $2.2\,(5.1)$. The resulting dilepton signal and the current LHC limits are shown in \Fig{fig-limits-dilepton} for the Higgs coupling of the model of \Sec{sec:fermionic} (left) and that of the model described in \App{app:SO5SO4} (right).

\begin{figure}[t]
\begin{center}
\includegraphics[width=0.55\textwidth]{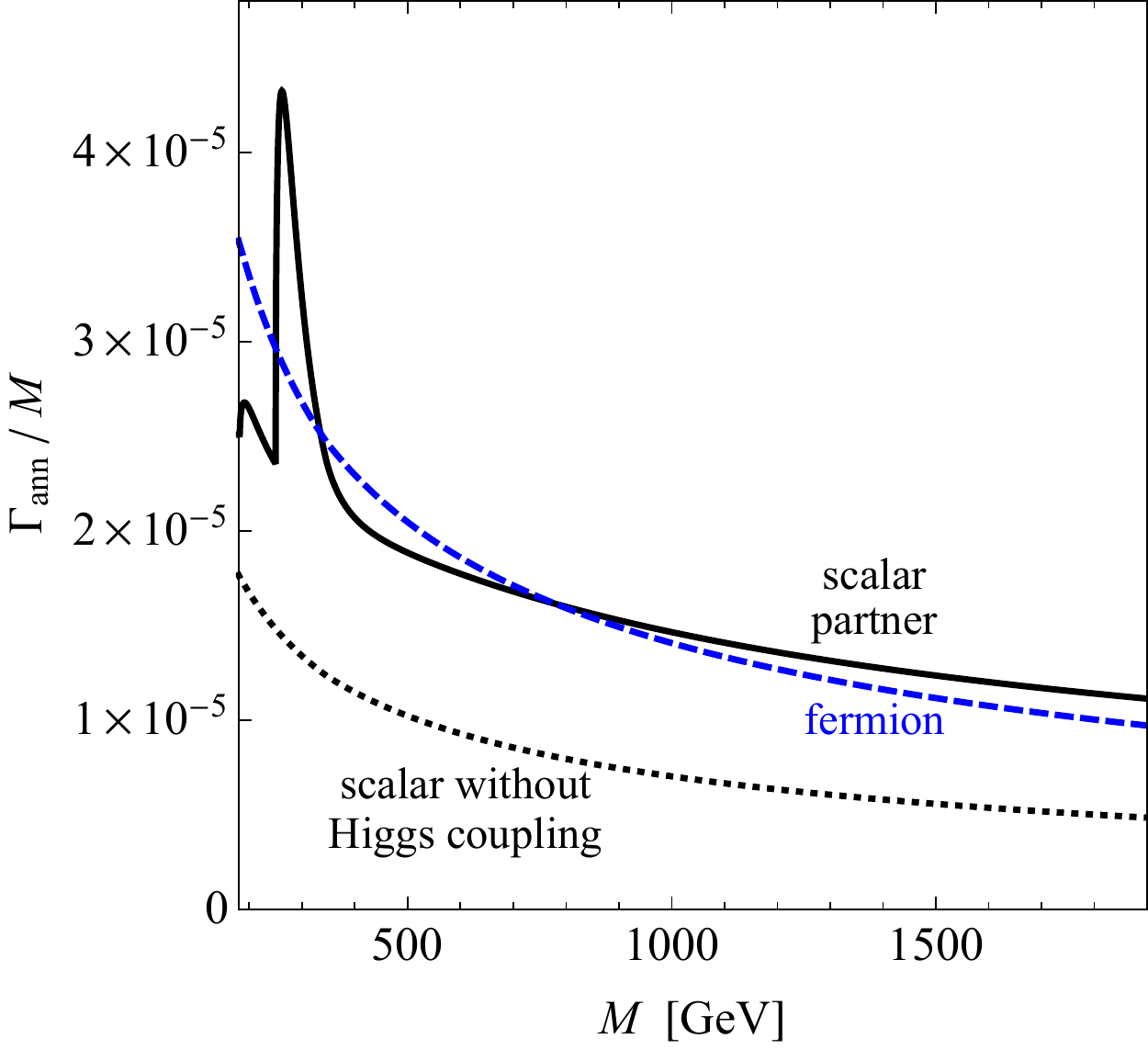}
\end{center}
\caption{Spin-0 bound state annihilation width as a function of the bound-state mass for $\SU(2)_L$-singlet constituents with electric charge $Q = -4/3$.  Shown are the cases of scalar top partners (solid black), scalars with no coupling to the Higgs (dotted black), and fermions (dashed blue), the latter of which is not affected by Higgs couplings.}
\label{fig:width}
\end{figure}

The bound-state annihilation signals computed above will  in general be diluted by the intrinsic decays of the constituent particles, unless the relative intrinsic width of the constituents, $\Gamma_{\rm decay}/m$, is much smaller than that corresponding to bound-state annihilation, $\Gamma_{\rm ann}/M$.  (A famous example where this condition is not satisfied is the SM top quark.)  While the constituent particle width for a two-body decay via a coupling $g$ is typically given by $\Gamma_{\rm decay}/m\sim g^2/16\pi^2 \sim 10^{-2} g^2$, the annihilation rate is inversely proportional to the cube of the Bohr radius such that $\Gamma_{\rm ann}/M \sim \alpha_{\rm ann}^2\alpha_s^3$, where $\alpha_{\rm ann}$ is the coupling responsible for the annihilation. Without the Higgs coupling, the annihilation width of the spin-0 $S$-wave bound states is dominated by the $gg$ contribution, i.e., $\alpha_{\rm ann}=\alpha_s$, which gives $\Gamma_{\rm ann}/M \sim \alpha_s^5 \sim 10^{-5}$. Annihilation modes enhanced by the Higgs coupling of top partners add a contribution of the same order of magnitude, in the case of scalar constituents only. As an example, \Fig{fig:width} shows the annihilation rates for the spin-0 bound states of $\SU(2)_L$ singlets with charge $-4/3$, showing the small expected value of $\Gamma_{\rm ann}/M \sim 10^{-5}$. (The bump in the plot occurs because annihilation into pairs of Higgses via the operator of Eq.~\eqref{eq:generic_spinzero} becomes kinematically allowed and then its rate quickly decreases due to a cancellation, as mentioned in the context of \Fig{fig-limits}.) For the spin-1 bound states (not shown), the width is even smaller because QCD-strength annihilation to either $gg$ or $q\bar q$ is absent.

Therefore, for the annihilation signals not to be diluted, the constituent intrinsic width should be somewhat suppressed. This can be the case in the presence of phase space suppression (either because only multi-body final states are possible or because one of the final-state particles is heavy) or if $g\ll1$ due to an approximate symmetry, as for the $\mathcal{Z}^T_2$ in our models. The top partners in our scenarios generically satisfy this condition, making the bound-state annihilation signals a rather model-independent experimental probe of these frameworks. Moreover, as can be seen in Fig.~\ref{fig-limits}, despite the smallness of the bound-state cross sections, meaningful limits are already being set in the mass range motivated by naturalness.

\subsection{Top partner pairs}
\label{subsec:pairprod}

We now turn to pair production of hypertwisted top partners at the LHC.
Given the strong bounds on stable colored particles, for example 1.2\,TeV for a color-triplet scalar with charge $2/3$~\cite{CMS-PAS-EXO-16-036,Aaboud:2016uth}, we assume that the top partners have an available decay mode, which requires their electric charge to differ from $2/3$ by an integer.
While the production cross section for colored top partners is large, their decays are model-dependent, depending on the specific way that the $\mathcal{Z}^T_2$ is broken. 
Various examples of pair-produced light colored scalars and fermions with exotic electric charges evading experimental constraints have been discussed recently in~\Refs{Kats:2016kuz,Blum:2016szr}.  Some of these (and other) decays can be realized in hypertwisted models.

\begin{figure}[t]
\begin{center}
\includegraphics[width=0.6\textwidth]{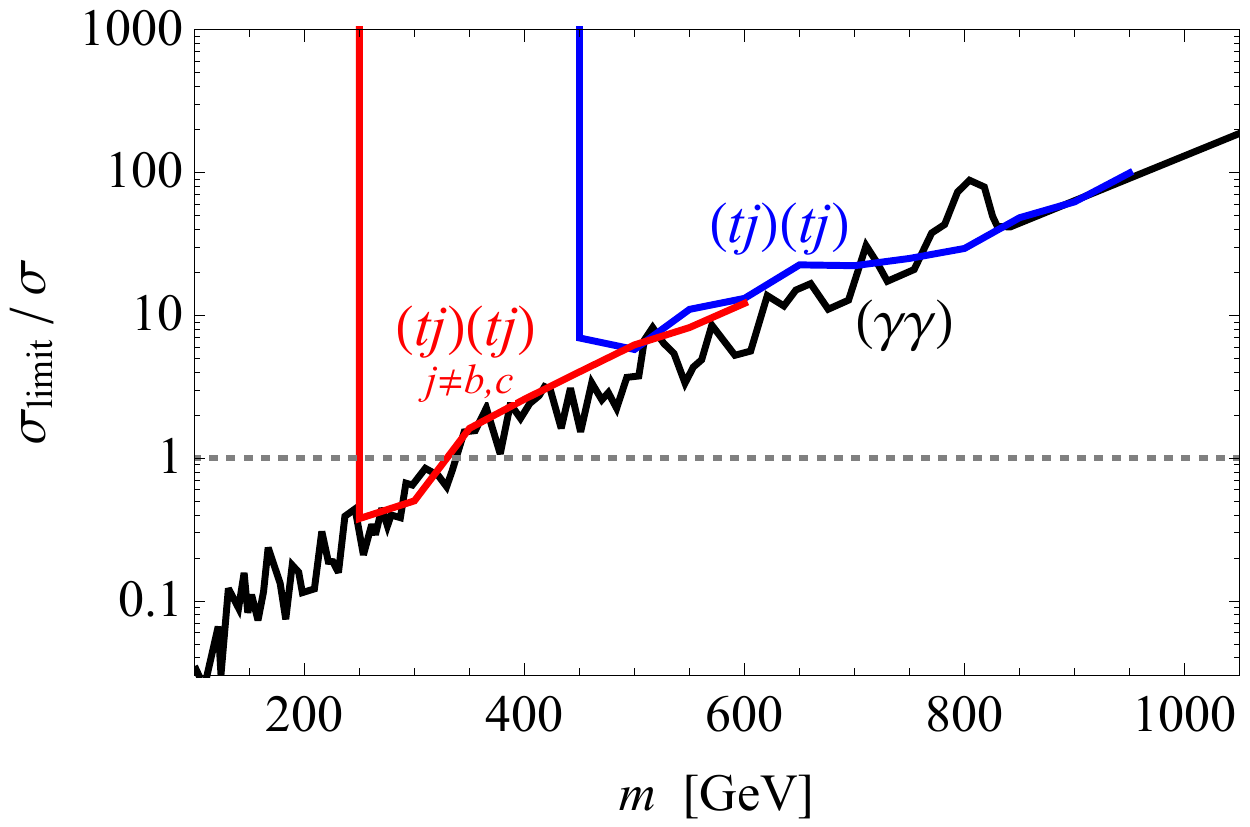}
\end{center}
\caption{Cross section limits on a color-triplet scalar with electric charge $-4/3$, as a function of its mass $m$.  Shown are CMS limits on top+jet decays based on~\Refs{CMS-PAS-B2G-12-008,Khachatryan:2016iqn} (red) and~\Ref{Chatrchyan:2013oba} (blue), using the 8~TeV dataset.  The limits from~\Refs{CMS-PAS-B2G-12-008,Khachatryan:2016iqn} (red) do not apply when the jet is a charm since the analysis employs loose $b$-tag vetoes.  Also shown is the limit on the bound state diphoton signal based on the ATLAS search~\cite{ATLAS-CONF-2016-059} (black), using 15.4~fb$^{-1}$ of the 13~TeV dataset. The limit from the analogous CMS search~\cite{CMS-PAS-EXO-16-027} (not shown) is similar.}
\label{fig:diphoton-vs-pairs}
\end{figure}

For example, in the hyperfolded SUSY framework presented in \Sec{sec:scalar}, a hyperfolded stop with charge $-4/3$ that is at least partially right-handed can decay (in the $\U(1)_{Y_F}$-broken phase) via the superpotential term\footnote{Note that a leptoquark-like coupling, which is entirely possible from the non-SUSY perspective for a particle with these quantum numbers, is incompatible with~\Eq{W-UUU} due to holomorphy, as it would require $W \supset \mb{U_F^c}^\dagger \mb{D^c} \mb{E^c}$.  That said, such terms could arise from the K\"ahler potential after SUSY breaking.} 
\beq
\mb{W} \supset \mb{U^c_F} \mb{U^c} \mb{U^c}
\label{W-UUU}
\eeq
as
\beq
\label{eq:Scalarjj}
\widetilde t_F \;\to\; \bar{t}\,\bar{c} \quad\mbox{or}\quad \bar{t}\,\bar{u} \,.
\eeq
These decays are almost unconstrained by the existing searches~\cite{CMS-PAS-B2G-12-008,Khachatryan:2016iqn,Chatrchyan:2013oba}, as shown in \Fig{fig:diphoton-vs-pairs}.
This channel reveals a particularly stark contrast between the case of hyperfolded stop squarks and the usual stop squarks.  In RPV  SUSY scenarios, squark decays to pairs of quarks may occur through the $\mb{U^c} \mb{D^c} \mb{D^c}$ superpotential operator.  However, this will only allow the stop to decay to two down-squarks, thus the top+jet final state is absent for stop decays in RPV SUSY.  For the hyperfolded stop of charge $-4/3$, however, this final state is allowed.  Thus searches for top+jet resonant pairs at the LHC probe a very interesting and unexplored region of SUSY-like models.

It is also interesting to consider the signatures of $\SU(2)_L$ partners of top partners, namely hyperfolded sbottoms. A sbottom with charge $-4/3$, for example, can also decay as in \Eq{eq:Scalarjj} via the operator
\beq
\mb{W} \supset \mb{D^c_F} \mb{U^c} \mb{U^c}
\label{W-DUU}
\eeq
in the presence of a right-handed component, subject to the same bounds as \Fig{fig:diphoton-vs-pairs}.   
Alternatively, the sbottom may decay via the operator
\beq
\mb{W} \supset (\mb{H_u} \mb{Q_F})(\mb{Q} \mb{Q})
\label{Y56-operator}
\eeq
(where parentheses enclose $\SU(2)_L$ singlets) as
\beq
\tilde b_F \;\to\; W^- \bar u \bar d \quad\mbox{or}\quad W^- \bar c \bar s \,.
\label{Wjj}
\eeq
We note that the 8\,TeV LHC limit on pair-produced particles decaying to $Wj$ is not very constraining~\cite{Aad:2015tba}. It is therefore plausible for the $Wjj$ decays of~\Eq{Wjj} to also be unobserved at this stage even for low masses.  A recent study~\cite{Blum:2016szr} found no constraints on this signature from re-interpretation of existing searches (intended to address completely different signatures) for masses above $240$~GeV.\footnote{For $\SU(2)_L$ doublets, constraints from electroweak precision tests need to be taken into account.  As shown in \Ref{Blum:2016szr}, these limits are not prohibitive.} It is plausible though that a dedicated search, if performed, will have some sensitivity. The left-handed stop, which is expected to be close in mass to the sbottom and have charge $-1/3$ in this scenario, can decay through the operator in \Eq{Y56-operator} into dijet pairs as $\tilde t_F \to \bar u \bar d$, which would be consistent with existing searches as long as it is heavier than about 500\,GeV~\cite{ATLAS-CONF-2016-084,CMS-PAS-EXO-16-029}.

We now turn to the hypertwisted composite Higgs scenarios discussed in \Sec{sec:fermionic} and \App{app:SO5SO4}. In standard composite Higgs scenarios, top partners typically decay to a $W$, $Z$, or $h$ boson and a top or bottom quark~\cite{DeSimone:2012fs}, including partners with exotic charges, which decay to $W^-b$ (for charge $-4/3$) or $W^+t$ (for charge $5/3$). Such decays can be seen to arise from the fermion Yukawa interactions.  In the models we consider here, however, the Yukawa interactions responsible for naturalness do not mediate such decays. As a result, other decays, depending on the details of the UV physics, may dominate.

For example, a fermionic top partner $T$ with charge $-4/3$ can decay
via the dimension-6 operator
\beq
{\cal L} \propto T^{c\dagger}_\alpha u^{c\dagger}_{i\,\beta} d_j^{c\,\alpha} d_k^{c\,\beta} + {\rm h.c.}
\eeq
(where $i,j,k$ are flavor indices and $\alpha,\beta$ are color indices) as
\beq
T \;\to\; jjj \quad\mbox{or}\quad \bar t jj \,.
\eeq
The constraints on $jjj$ decays are not yet prohibitive as long as the jets do not include $b$ jets~\cite{Chatrchyan:2013gia}, and there are no dedicated searches for pairs of $tjj$ resonances. Another interesting example, again for a $T$ with charge $-4/3$, is the operator
\beq
{\cal L} \propto \epsilon_{\alpha\beta\gamma}\,T^{c\alpha\dagger}  q^\beta_i q^\gamma_j e^c_k + {\rm h.c.}
\eeq
mediating the decay
\beq
T \;\to\; \tau^- jj \,,
\eeq
which might be underconstrained. There do exist, however, relatively strong limits on the somewhat similar signature with particles decaying to $\tau j$ from the CMS search~\cite{Sirunyan:2017yrk}.

As a final note, in the case of an exact $\mathcal{Z}^T_2$ symmetry the top partner could decay to a neutral $\mathcal{Z}^T_2$-odd particle which can be a dark matter candidate.  However, missing energy searches (e.g.~\cite{Aad:2015iea,Aad:2015pfx,Aaboud:2016zdn,CMS-PAS-SUS-16-004,Sirunyan:2016jpr,CMS-PAS-SUS-16-033,CMS-PAS-SUS-16-049}) generically set strong bounds on that possibility.

\section{Summary and outlook}
\label{sec:summary}

The lack of experimental clues for an extension to the SM does not imply that the electroweak hierarchy problem has gone away; rather, the puzzle of weak-scale naturalness is now more acute than ever.
Already for some time, it has been necessary to reconsider the basic assumptions about weak-scale naturalness and its associated phenomenology.
For example, it has recently been proposed, in radical departures from common approaches, that perhaps the underlying explanation for the hierarchy between the electroweak and Planck scales is not that it has been stabilized in the quantum theory by an underlying symmetry, but rather that it emerges as a result of cosmological dynamics or vacuum selection effects~\cite{Graham:2015cka,Arkani-Hamed:2016rle,Arvanitaki:2016xds}.

While these radical departures must be taken seriously, it is still possible that reality may be more akin to conventionally-considered naturalness scenarios, with a spectrum of partner particle states within reach of the LHC.
Even within these more conventional scenarios, though, there can be dramatic departures in the expected experimental signatures through relatively minor tweaks to the underlying symmetry structures.
This is best seen in neutral naturalness scenarios~\cite{Chacko:2005pe,Chacko:2005vw,Chacko:2005un,Falkowski:2006qq,Chang:2006ra,Batra:2008jy,Craig:2013fga,Craig:2014aea,Craig:2014roa,Geller:2014kta,Barbieri:2015lqa,Low:2015nqa,Craig:2015pha,Craig:2016kue,Barbieri:2016zxn,Katz:2016wtw,Contino:2017moj,Badziak:2017syq,Cai:2008au,Burdman:2006tz,Cohen:2015gaa,Poland:2008ev,Batell:2015aha}, where top partners are inert under $\SU(3)_C$ and thereby immune to the most stringent bounds on naturalness from the LHC.
Thus, the experimental implications of weak-scale searches are highly sensitive to the detailed mechanism for how naturalness is achieved in the UV.

In this work, we introduced the possibility of colorful twisted top partners, which still carry $\SU(3)_C$ but have exotic electric charges, and we showed how hypertwisted scenarios could be embedded in consistent UV structures.
From the perspective of electroweak naturalness, the electric charges of weak-scale top partners are largely irrelevant, since the one-loop cancellation of the leading top quark divergence persists for any charge assignment.
From the perspective of collider phenomenology, though, electric charges have a huge impact on the allowed decay modes of the top partners, even resulting in stable colored particles in the most exotic cases.
So while the direct searches for ordinary top partners at the LHC may lead to the impression that colored top partners are close to extinction in the best-motivated mass ranges, hypertwisted top partners can still be viable due to their exotic decay phenomenology.

The most model-independent prediction of hypertwisted scenarios is the presence of partnerium bound states.
Similar to (but more robustly predicted than) stoponium, top-partnerium can be produced through gluon fusion and annihilate to pairs of photons or electroweak bosons, such that searches for narrow diboson resonances will play an important role in constraining this rich class of scenarios.
There are also more model-dependent possibilities that arise in hypertwisted scenarios.
Search channels that are not usually associated with conventional top partners---such as pair production of top-plus-jet resonances in the scalar partner case or multibody decays from nonrenormalizable operators in the fermionic partner case---are crucial for covering the natural parameter space in hypertwisted models. 
Combined with the enormous datasets still to be accumulated by the LHC experiments, we hope these searches help expand the experimental frontier of weak-scale naturalness at the LHC.

\acknowledgements{
We thank Nathaniel Craig, Diptimoy Ghosh, and David Pinner for useful discussions, and Andrea Wulzer for comments on the draft.
Feynman diagrams in this paper were generated with TikZ-Feynman~\cite{Ellis:2016jkw}, and we thank Joshua Ellis to for his support using the package.
The work of GP is supported by grants from the BSF, ERC, ISF, Minerva, and the Weizmann-UK Making Connections Programme.
The work of YS and JT is supported by the U.S. Department of Energy (DOE) under grant contract numbers DE-SC-00012567 and DE-SC-00015476.
}

\appendix

\section{Hypertwisted composite Higgs with custodial protection}
\label{app:SO5SO4}

In this appendix, we show a hypertwisted version of the $\SO(5)/\SO(4)$ minimal composite Higgs model~\cite{Agashe:2004rs}. The advantage of this model is that it has custodial protection, which relaxes the tension with electroweak precision tests. 
Below, we see that the cancelation of the top loop contribution to the Higgs mass is similar to the model of Sec.~IV~B of~\Ref{Batell:2015aha} and requires three top partners. 

We start with a global symmetry
\be
\SO(5)_G \times \SU(2)_F \times \U(1)_Z \,,
\ee
where $\SU(2)_F$ plays the same role as in \Sec{sec:fermionic}.  We introduce a (linear) sigma field that transforms under $(\SO(5)_G,\SU(2)_F)_{\U(1)_Z}$ as
\be
(\boldsymbol{5},\boldsymbol{1})_{0} : \quad \Phi \,,
\ee
which spontaneously breaks
\be
\SO(5)_G \to \SO(4) \,,
\ee
leaving $\SU(2)_F \times \U(1)_Z$ unaffected.

We can expand the field $\Phi$ as 
\begin{align}
	\label{eq:Phi54}
	\Phi = \exp\left[ i \sqrt{2}\frac{h^a T^a_G}{f} \right]	\begin{pmatrix}
		0 \\ 0 \\ 0 \\ 0 \\ f
	\end{pmatrix}
	=
	\frac{\sin(|h|/f)}{|h|/f}
	\begin{pmatrix}
		h_1 \\ h_2 \\ h_3 \\ h_4 \\ |h|\cot(|h|/f)
	\end{pmatrix}
	\quad
	\overset{h=\langle h \rangle}{\Longrightarrow}
	\quad
	f
	\begin{pmatrix}
		0 \\ 0 \\ s_\epsilon \\ 0 \\ c_\epsilon
	\end{pmatrix} \, ,
\end{align}
where $T^a_G$ with $a=1,\ldots,4$ are the broken generators (for the algebra, see e.g.~\cite{Agashe:2004rs}), $|h|= \sqrt{h_1^2+h^2_2+h_3^2+h^2_4}=\sqrt{2H^\dagger H}$, and $s_\epsilon$ is the sine of $\epsilon \equiv v/f$, where $v \approx 246$\,GeV is the Higgs vev. The SM electroweak gauge group, $\SU(2)_L\times \U(1)_Y$, is identified with the following generators which are weakly gauged:
\begin{align}
	T_L^{1,2,3} = T_{G,L}^{1,2,3}\,, \qquad\qquad
	Y = T^3_{G,R} + Z + \left(\frac{2}{3} - y_T\right)T^3_F \,,
\end{align}
where $T_{G,L/R}^{1,2,3}$ are the generators of the $\SU(2)$ factors in $\SO(4)\simeq \SU(2)_L\times \SU(2)_R$.

The relevant matter content in the top sector is
\begin{align}
	\label{eq:Q54}
	(\boldsymbol{\bar 5},\boldsymbol{2})_{\frac{y_T}{2}+\frac{1}{3}} : \quad
Q =	\frac{1}{\sqrt2}\begin{pmatrix}
	b & \widetilde{q}_u \\
	-ib & i\widetilde{q}_u \\
	t & \widetilde{q}_d \\
	it & -i\widetilde{q}_d \\
	0 & \sqrt{2}\widetilde{T}
	\end{pmatrix} \, , \qquad\quad
	 (\boldsymbol{1},\boldsymbol{\bar 2})_{-\frac{y_T}{2}-\frac{1}{3}} : \quad
Q^c= 	-\begin{pmatrix}
	t^c & \widetilde{T}^c
	\end{pmatrix} \,,
\end{align}
where $q \equiv (t,b)$ and $\widetilde{q} \equiv (\widetilde{q}_u,\widetilde{q}_d)$ are $\SU(2)_L$ doublets. In the above, $Q$ contains a complete pseudo-$\SO(5)_G$ multiplet split between both columns.  This can be obtained by starting with two complete $\SO(5)_G$ multiplets in the doublet (i.e.\ a full $\SO(5)_G \times \SU(2)_F$ bifundamental) and decoupling a pseudo-$\SO(5)_G$ multiplet that is split across both columns, in analogy to the primed fields in \Eq{eq:QQccomp} of \Sec{sec:fermionic}.

The SM charges of the fields in this model are given in \Tab{tab:CompositeSO5SO4} (while the decoupled states have electric charges $2/3$, $5/3$, $y_T$ and $y_T+1$).  The Yukawa interaction is as in \Eq{eq:LY} and an expansion in $H/f$ leads to
\begin{align}
	\label{eq:LYexpand54}
	\cL_{Y} \supset
	-\lambda_t q H t^c - \lambda_t \left(f - \frac{H^\dagger H}{f}\right) \widetilde{T} \widetilde{T}^c 
	- \lambda_t  \widetilde{q} H^\dagger \widetilde{T}^c - M_{\widetilde{q}} \widetilde{q} \widetilde{q}^{c} 
	 + \cO(1/f^2) \, ,
\end{align}
where $M_{\widetilde{q}}$ is a vector-like mass term, added to ensure that $\widetilde{q}$ is massive. Note that the $H^\dagger H \widetilde{T} \widetilde{T}^c$ interaction is a factor of 2 larger than in \Eq{eq:LYexpand}, which has an impact on the phenomenology of the spin-1 partnerium, as seen in \Fig{fig-limits-dilepton}.  The $\widetilde{q} H^\dagger \widetilde{T}^c$ term is crucial for completing the cancellation of the top loop, so the members of the $\widetilde q$ doublet are top partners as well. However, a study of their phenomenology is outside the scope of this work.

\begin{table}[t]
\begin{center}
\begin{tabular}{c@{$\quad$}c@{$\quad$}c@{$\quad$}c}
\hline
\hline
  & $\SU(3)_C$ & $\SU(2)_L$ & $\U(1)_Y$ \\
  \hline \hline
 $H$   & $\boldsymbol{1}$         & $\boldsymbol{2}$ & $1/2$ \\ \hline
 $q$   & $\boldsymbol{3}$         & $\boldsymbol{2}$ & $1/6$ \\
 $t^c$ & $\boldsymbol{\bar3}$   & $\boldsymbol{1}$ & $-2/3$ \\\hline
 $\widetilde{T}$   & $\boldsymbol{3}$         & $\boldsymbol{1}$ & $y_T$  \\
 $\widetilde{T}^c$ & $\boldsymbol{\bar3}$   & $\boldsymbol{1}$ & $-y_T$ \\
 $\widetilde{q}$  & $\boldsymbol{3}$         & $\boldsymbol{2}$ & $y_T + 1/2$ \\
 $\widetilde{q}^c$ & $\boldsymbol{\bar3}$ & $\boldsymbol{2}$ & $-(y_T + 1/2)$ \\
 \hline
 \hline
\end{tabular}
\caption{The SM quantum numbers of the different fields of the hypertwisted $\SO(5)/\SO(4)$ composite Higgs model.}
\label{tab:CompositeSO5SO4}
\end{center}
\end{table}%

The fermion mass terms, based on \Eqs{eq:Q54}{eq:LYexpand54}, are
\begin{align}
	\cL_{\rm mass}
=	-M_{2/3}\, tt^c - \begin{pmatrix} \widetilde{T} & \widetilde{q}_d \end{pmatrix} M_{y_T} \begin{pmatrix} \widetilde{T}^c \\ \widetilde{q}^{c}_d \end{pmatrix}
- M_{y_T+1}\, \widetilde{q}_u \widetilde{q}^{c}_u
+ \mbox{h.c.} \,,
\end{align}
where
\begin{align}
M_{2/3} = \lambda_t f \frac{s_\epsilon}{\sqrt{2}} = m_t  \, , \quad\quad 
M_{y_T} 
=	\begin{pmatrix}
\lambda_t f c_\epsilon & 0 \\
\lambda_t f s_\epsilon/\sqrt2 & M_{\widetilde{q}}
\end{pmatrix}, \quad\quad
M_{y_T+1} = M_{\widetilde{q}} \, .
\end{align}
The partner masses are $\sim\lambda_t f c_\epsilon$ and $\sim M_{\widetilde{q}}$, where either one of them could be the lightest.
It is straightforward to verify that ${\rm tr}[M_{2/3}^\dagger M_{2/3} \Lambda^2]+{\rm tr}[M_{y_T}^\dagger M_{y_T} \Lambda^2]+{\rm tr}[M_{y_T+1}^\dagger M_{y_T+1} \Lambda^2]$ is independent of $s_\epsilon$ and the model is free of quadratic divergences at one loop.  Finally, we note that the discussion regarding the decays of the top partner is similar to \Secs{sec:ColorTwist}{subsec:pairprod}.

\section{Impact of the Higgs on bound state production}
\label{impact-on-production}

In this appendix, we show that Higgs boson exchange has a negligible effect on partnerium bound-state production in the parameter range of interest.

In the scalar top partner case, for both right- and left-handed hyperfolded stops, the interaction in \Eq{eq:LSUSY} produces a ``higgs force'' coupling of the form
\beq
{\cal L} \supset -\kappa\, v\, h \,\tilde t_F^\ast\, \tilde t_F \,,
\eeq
with $\kappa = \lambda_t^2 \approx 1.0$. In the nonrelativistic limit, this gives rise to the Yukawa potential
\beq
V_h(r) = -\frac{\alpha_h}{r}\, \exp(-m_h r) \,,
\label{higgs-potential}
\eeq
where (see e.g.~\cite{Efimov:1999tx,Darewych:2000,Petraki:2015hla})
\beq
\alpha_h = \frac{\kappa^2}{16\pi}\frac{v^2}{m^2} \approx 4.7 \times 10^{-3}\left(\frac{\kappa}{\lambda_t^2}\right)^2\left(\frac{500~{\rm GeV}}{m}\right)^2 \,,
\label{alpha_h}
\eeq
where $m$ is the stop mass.

In the fermionic top partner case, the interactions of \Eq{eq:LYexpand} produce a coupling of the form
\beq
{\cal L} \supset \frac{\kappa v}{2m}\, h\, T\, T^c  + \text{h.c.} \,,
\eeq
which leads to the same result as in Eqs.~\eqref{higgs-potential}--\eqref{alpha_h} (see, e.g.,~\cite{Strassler:1990nw}). With \Eq{eq:LYexpand} itself,
\beq
\kappa = \frac{2m_t m}{v^2}\,\arctan\left(\frac{m_t}{m}\right) \,,
\eeq
which reduces to the SM $\lambda_t^2$ for $m \gg m_t$ (i.e.\ $f \gg v$), while for the model described in \App{app:SO5SO4}, \Eq{eq:LYexpand54} gives
\beq
\kappa = \frac{2\sqrt2\,m_tm}{v^2}\,\arctan\left(\frac{\sqrt2\,m_t}{m}\right) \,,
\eeq
which reduces to $2\lambda_t^2$ in the same limit.

We estimate that in the range of parameters of interest, the physics of the bound state in the combined QCD and Higgs potential,
\beq
V(r) = - \frac{C_{\mathbf 3}\bar\alpha_s}{r} - \frac{\alpha_h}{r}\, \exp(-m_h r)
\eeq
(where $C_{\mathbf 3} = 4/3$, and $\bar\alpha_s \approx 0.14$, as it is evaluated at the scale of the bound state), remains dominated by the QCD interaction. Suppose, for instance, we neglect the (very significant) exponential suppression and are left with the Coulomb potential
\beq
V_0(r) = -\frac{\alpha_{\rm eff}}{r}
\eeq
with
\beq
\alpha_{\rm eff} = C_{\mathbf 3}\bar\alpha_s + \alpha_h \approx C_{\mathbf 3}\bar\alpha_s\left[1 + 0.025\left(\frac{\kappa}{\lambda_t^2}\right)^2\left(\frac{500~{\rm GeV}}{m}\right)^2\right] .
\eeq
Even in this limit, the bound state production cross section, which is proportional to
\beq
|\psi(\mathbf{0})|^2 \propto \alpha_{\rm eff}^3 \propto 1 + 0.076\left(\frac{\kappa}{\lambda_t^2}\right)^2\left(\frac{500~{\rm GeV}}{m}\right)^2
\eeq
is enhanced only by roughly $50\%$ for $m = 200$~GeV and $2\%$ for $m = 1$~TeV, for $\kappa = \lambda_t^2$. In reality, the exponential suppression due to the Higgs mass makes this enhancement significantly smaller. Note, for instance, that in the Coulomb approximation, the RMS size of the bound state is $r_{\rm RMS} = 3\sqrt 3/(2\alpha_{\rm eff}m) \approx 13/m$, which is larger than $1/m_h$ in the mass range of interest ($m \lesssim 1$~TeV). Alternatively, note that the condition for a bound state to even exist in a Higgs-only potential~\cite{Rojers:1970},
\beq
D \equiv \frac{\alpha_{\rm eff}m}{2m_h} \gtrsim 0.84 \,,
\eeq
would only be satisfied for $m \gtrsim 1$~TeV even if both the Higgs and the gluon contributions were included in $\alpha_{\rm eff}$.

Considering that the bound-state rates are in any case subject to an uncertainty of roughly a factor of $2$~\cite{Kats:2016kuz}, we neglect the effects of Higgs exchange on the binding in the current work.

\bibliographystyle{utphys}
\bibliography{CTTP_bib}

\end{document}